 \documentclass[final,5p,times,twocolumn,authoryear]{elsarticle}

\usepackage{amssymb}
\usepackage{lipsum}
\newcommand\arcdeg{\mbox{$^\circ$}}
\newcommand{\rs}{$\rm R_{\odot}$}
\usepackage{siunitx}

\journal{PASJ}

\begin{document}

\begin{frontmatter}



\title{X-ray/H$\mathrm{\alpha}$ scaling relationships in stellar flares}


\author[1]{Hiroki K{\small{AWAI}}$^{\ast}$}
\author[1]{Yohko T{\small{SUBOI}}$^{\ast}$}
\affiliation[1]{organization={Department of Physics, Chuo University, 1-13-27 Kasuga, Bunkyo-ku, Tokyo 112-8551, Japan}}

\author[1,2]{Wataru B. I{\small{WAKIRI}}}
\affiliation[2]{organization={RIKEN, 2-1 Hirosawa, Wako, Saitama 351-0198, Japan}}

\author[3]{Yoshitomo M{\small{AEDA}}}
\affiliation[3]{organization={Institute of Space and Astronautical Science, Japan Aerospace Exploration Agency, 3-1-1 Yoshinodai, Chuo-ku, Sagamihara, Kanagawa 252-5258, Japan}}

\author[4]{Satoru K{\small{ATSUDA}}}
\affiliation[4]{organization={Graduate School of Science and Engineering, Saitama University, 255 Shimo-Okubo,Sakura, Saitama, 338-8570, Japan}}

\author[1]{Ryo S{\small{ASAKI}}}
\author[1]{Junya K{\small{OHARA}}}

\author{the MAXI TEAM}

\begin{abstract}
% Text of abstract
We report on the results of our simultaneous observations of three large stellar flares with soft X-rays (SXRs) and an H$\mathrm{\alpha}$ emission line from two binary systems of RS CVn type. The energies released in the X-ray and H$\mathrm{\alpha}$ emissions during the flares were $10^{36}$--$10^{38}$ and $10^{35}$--$10^{37}$ erg, respectively. This renders the set of the observations as the first successful simultaneous X-ray/H$\mathrm{\alpha}$ observations of the stellar flares with energies above $10^{35}$ erg; 
although the coverage of the H$\mathrm{\alpha}$ observations of the stellar flares with energies above $10^{35}$ erg; although the coverage of the H$\mathrm{\alpha}$ observations was limited, with $\sim$10\% of the $e$-folding time in the decay phase of the flares, that of the
SXR ones was complete. 
Combining the obtained physical parameters and those in literature for solar and stellar flares, we obtained a good proportional relation between the emitted energies of X-ray and H$\mathrm{\alpha}$ emissions for a flare energy range of $10^{29}$--$10^{38}$ erg. The ratio of the H$\mathrm{\alpha}$-line to bolometric X-ray emissions was $\sim$0.1, where the latter was estimated by converting the observed SXR emission to that in the 0.1--100 keV band according to the best-fitting thin thermal model.
We also found that the $e$-folding times of the SXR and H$\mathrm{\alpha}$ light curves in the decaying phase of a flare are in agreement for a time range of $1$--$10^4$~s.
Even very large stellar flares with energies of six orders of magnitude larger than the most energetic solar flares follow the same scaling relationships with 
solar and much less energetic stellar flares. This fact suggests that their physical parameters can be estimated on the basis of the known physics of solar and stellar flares.
\end{abstract}

\begin{keyword}
%% keywords here, in the form: keyword \sep keyword, up to a maximum of 6 keywords
stars: activity  --- stars: flare --- X-rays: stars \\

Accepted by PASJ 2022 January 26 \\
$^{\ast}$E-mail: kawai@phys.chuo-u.ac.jp; tsuboi@phys.chuo-u.ac.jp

\end{keyword}

\end{frontmatter}

%\tableofcontents



\section{Introduction} \label{sec:int}
\label{introduction}

Large stellar flares with energies between 10$^{34}$--10$^{39}$~erg have been detected \citep{Tsuboi:2016,Sasaki:2021} with the all-sky X-ray monitor, Monitor of All-sky X-ray Image \citep[MAXI][]{Matsuoka:2009} since its launch in 2009. The majority of the detected flares are estimated to have loop lengths likely larger than the radii of their host stars by a factor of $\sim$5. 

The MAXI-detected flares have extremely large energy-related parameters: luminosities of $10^{31-34}$ erg s$^{-1}$ in the 2--20 keV band, emission measures of 10$^{54-57}$ cm$^{-3}$, $e$-folding times of 1 hr to 1.5 d, and total radiative energies of 10$^{34-39}$ erg. \citet{Tsuboi:2016} found a universal correlation between the flare duration and peak X-ray luminosity, combining the X-ray flare data of nearby stars and the Sun. Moreover, they found that the MAXI-detected flares extended the established correlation between the flare-peak emission measure and temperature for solar flares and small stellar ones \citep{Shibata:1999}. 
Given that the correlations hold over a broad range of energies from solar micro flares to large stellar flares, the correlation suggests the presence of some common mechanism governing flare-loop formation and its cooling processes.

One of the other important observational probes for flare physics is the H$\mathrm{\alpha}$ emission. In the solar flare, H$\mathrm{\alpha}$ is known to be emitted from the foot of the loop, called the ``H$\mathrm{\alpha}$ Ribbon'', and/or a loop located at the inner side of a soft X-ray (hereafter, SXR) flare loop, called a postflare loop \citep[see figure 42 in ][]{Shibata:2011}.  Since the emission region and mechanism are different from those of the SXR, the information obtained from the H$\mathrm{\alpha}$ emission is complementary to that from the SXR emission. Hence, the study of the relationship between the SXR and H$\mathrm{\alpha}$ flares from distant stars, where imaging observations are difficult unlike the Sun, can provide key information about the overall geometrical structure and evolution mechanism of stellar flares.

\citet{Butler:1988} and \citet{Butler:1993} derived a positive relation between the emitted energies with the H$\mathrm{\gamma}$ emission line and SXR.
\citet{Veronig:2002} performed a detailed study of solar flares and reported on a positive relation between the decay times ($e$-folding times) of the observed fluxes of the H$\mathrm{\alpha}$ line and SXR.
However, no studies have ever been conducted about samples containing very large stellar flares, the energy of which can reach orders of magnitude larger than that of the largest solar flares, such as those detected with MAXI. Consequently, whereas physics of solar flares is comparatively well understood, our understanding of flare physics of much larger or more active stars than the Sun remains poor and not much more than speculation.

Here we report on the results of our H$\mathrm{\alpha}$ observations (see section~\ref{sec:obs} for the observation details) conducted immediately after after triggers of MAXI detection of large stellar flares with emitted energies of up to $10^{38}$ erg, combining the simultaneous observations with MAXI (section~\ref{sec:obs}) and continuous H$\mathrm{\alpha}$ monitoring. Specifically, we focus on three flares with the best statistics and analyze them via the methods described in section~\ref{sec:res}.
These are the first simultaneous observational samples ever reported of the H$\mathrm{\alpha}$ and SXR emissions of stellar flares with energies over $10^{35}$ erg, which is the previous record \citep[a flare from II Peg reported by][]{Butler:1993}.
Then we investigate the relations of the energy and $e$-folding time between the H$\mathrm{\alpha}$ and SXR emissions and discuss them (section~\ref{sec:dis}) before summarizing our result in section~\ref{sec:sum}.

\section{Observations} \label{sec:obs}
\subsection{MAXI} \label{subsec:maxi}
SXR is taken with the Monitor of All-sky X-ray Image, MAXI \citep{Matsuoka:2009} . 
MAXI is an astronomical X-ray observatory mounted on the International Space Station (ISS). In this analysis, we used data from the Gas Slit Camera \citep{Mihara:2011}, which is sensitive in the 2--30~keV band. It consists of 12 proportional counters, each of which employs carbon-wire anodes to provide one-dimensional position sensitivity. A pair of counters form a single camera unit; hence the instrument consists of six camera units. The six camera units are assembled into two groups whose fields of view (FoVs) are pointed toward the tangential direction of the ISS\ motion along the earth horizon and the zenith direction. The FoVs are 160$\arcdeg$~$\times$~3$\arcdeg$, which corresponds to 2\% of the whole sky. These counters are not operated in the regions with high particle background, such as the South Atlantic Anomaly and at absolute latitudes higher than $\sim$~40$\arcdeg$, or in the vicinity of the Sun (within $\sim$~5$\arcdeg$).
Hence the Gas Slit Camera has an operating duty ratio of $\sim$~40\% and scans about 85\% of the whole sky per orbit of the ISS.
The time interval of MAXI observations is $\sim$92 min, which is the same as the orbital period of the ISS. The exposure time of a target is $\sim$50 s per orbit.

\subsection{SCAT}\label{subsec:nicer}

\begin{figure}
    \centering
    \includegraphics[width=8.5cm, angle=0]{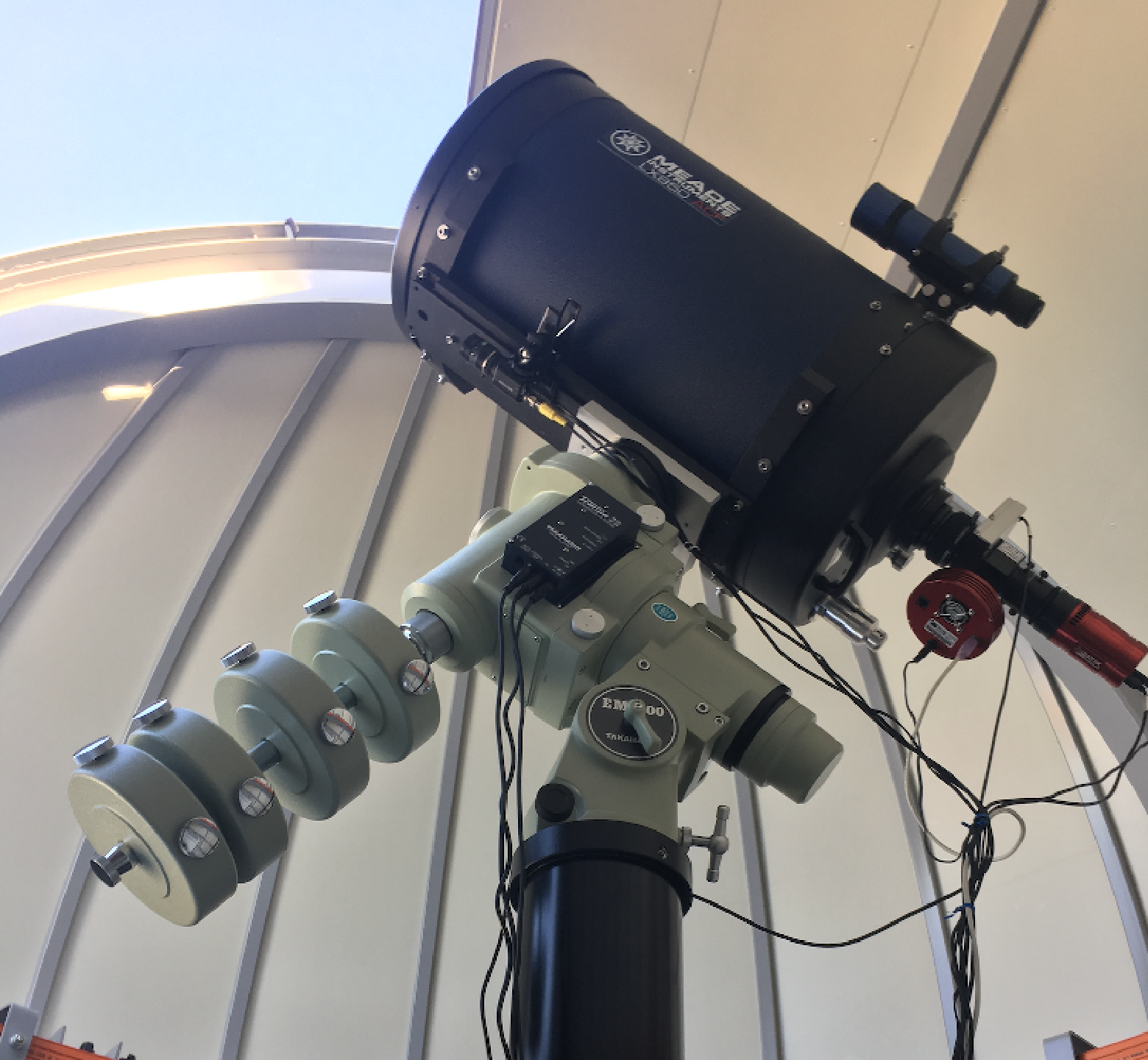}\\
    \caption{
    SCAT and its basic performance. Mirror--MEADE F8ACF35 (356 mm$\phi$, FL  2845 mm). Equatorial mount--Takahashi EM-400 Temma2Z. Finder--WAT-910HX/RC, CBCM5018-MP2 (50 mm). Grating--Shelyak Alpy 600 (Resolution power $R =$ 600). Sensor--ATIK 460EX. Sensitivity--10 mag for 600s exposure. Band width--3700--7400 $\textrm{\AA}$.
    }
    \label{fig:scat}
\end{figure}

The Spectroscopic Chuo-university Astronomical Telescope (SCAT; figure \ref{fig:scat}) is a MEADE 36 cm diameter optical telescope built on the rooftop of a building on the campus of Chuo University, situated in downtown of Tokyo, Japan (latitude = $\ang{35;42;30}$N, longitude = $\ang{139;44;54}$E).
SCAT is equipped with an ATIK 460EX CCD camera with a Shelyak Alpy 600 spectrometer, suitable for low-resolution spectroscopic observations of the H$\mathrm{\alpha}$ emission line. The spectrometer covers wavelengths of 3700 to 7500 $\textrm{\AA}$ with a spectral resolution of $R=600$. SCAT has the limiting magnitude at 6000~$\textrm{\AA}$ of about 10 mag. Its signal-to-noise ratio is 10 for an exposure of 600 s. 
SCAT observes the sky when the elevation is larger than $30\arcdeg$. This means that an area of the sky with a declination angle of larger than $-20\arcdeg$ is observable.

\subsection{Simultaneous observations}\label{subsec:simul}

\begin{table}[htbp]
    \centering
	\caption{
	{Time history of our simultaneous observations}
	}
	\scriptsize
	\renewcommand\arraystretch{2}
	{\fontsize{9pt}{6pt}\selectfont
	\begin{tabular}{lcccccccc} \hline
	    &&&Start time of the \\
        & Nova alert & SXR peak time & SCAT observation \\
		Flare ID & (MJD) & (MJD) & (MJD)   \\ \hline
		Flare 1 & 57714.1651 & 57714.2612 & 57714.4163 \\
	    Flare 2 & 58446.7929 & 58446.7587 & 58447.4228 \\
	    Flare 3 & 58157.9849 & 58158.0790 & 58158.3802 \\
		  \hline
	\end{tabular} 
	}
	\label{tb:flare_times}
\end{table}

{
We made use of ``nova alerts'' \citep{Negoro:2016} to do follow-up observations of MAXI-detected stellar flares. 
A time history of the nova alerts of three selected flares along with the SXR flare peak and the start time of the H$\mathrm{\alpha}$ emission is shown in Table \ref{tb:flare_times}. The flare peaks of Flares 1, 2, and 3 appeared about $+0.10$, $-0.04$, and $-0.09$ d, respectively, after the corresponding nova alert.
}
{
This automated alert system identifies X-ray transients from the image taken in real time, and classifies them into four types of events: ``Burst'', ``Alert'', ``Warning'', and ``Info". While ``Burst'' is automatically reported to scientists across the world via the ``MAXI Mailing List (MAXI ML)'', ``Alert'' and ``Warning'' are examined by duty scientists before a notification is circulated to MAXI ML. We started follow-up observations with SCAT, immediately after receiving a signal from ``nova alerts'' as duty scientists of MAXI. }
{
The detections of all the MAXI flares which we work on or mention in this paper (seven flares) were circulated to the MAXI ML. 

}

Apart from the flare periods, SCAT regularly monitored the sources, and the data were used to identify their characteristics during the persistent phase. The sources were selected from the MAXI-detected flare sources
(e.g., \citeauthor{Sasaki:2017} 2017).

During the period between 2016 November and 2021 October, we successfully made SCAT observations of seven stellar flares to follow up MAXI-alerted events. Out of the seven flares, we selected three flares for our analysis presented in this paper via the following criteria. First, the X-ray data of the flare must have sufficient statistics to determine the $e$-folding time  (accordingly, Flare~5 was excluded; see Appendix~\ref{sec_appendix}). Secondly, the decay phase of the flare must have been observed with SCAT (Flares~6 and 7 thus excluded)\footnote{We did make H$\mathrm{\alpha}$ observations during the period at around their flare peaks but not in their decay phases due to bad weather. 
}. Thirdly and lastly, Flare~4, which showed clear multiple peaks in the SCAT light-curve, was excluded because our analysis relies on a simple-profile light curve in the decaying phase.
The three selected flares are referred to as Flares 1, 2, and 3. Flares 1 and 2 occurred at UX Ari whereas Flare 3 did so at AR Psc. The detection history is listed in table {\ref{tb:flare_times}}.

\subsection{Objects}\label{subsec:objects}

\subsubsection{UX Ari}\label{subsec:uxari}

UX Arietis (UX Ari, HD 21242, HIP 16042) is a binary system of RS CVn type \citep{Fekel:1986}, located at a distance of 50.2~pc \citep{ESA:1997}.  It consists of a primary K0 subgiant and a secondary G5 main-sequence star.
The synchronous rotation period is reported to be 6.43791~$\pm$~0.00006 d \citep{Carlos:1971}. 
The CHARA six-telescope optical long-baseline array revealed that its spots can fill about 62\% of the primary surface \citep{Hummel:2017}.
{\cite{Hummel:2017} reported that the radii of the primary and secondary stars are 5.6~$\pm$~0.1~\rs and 1.6~$\pm$~0.2~\rs, respectively, with an inclination angle of 125$\arcdeg$.0~$\pm$~0$\arcdeg$.5 and a separation of 18.8~$\pm$~0.1~\rs ~(table~\ref{tb:uxari}). }

UX Ari has been known as a very active flare-star binary \citep{Feldman:1978,Tsuru:1989,Massi:1998,Massi:2002,Catalano:2003,Massi:2005,Peterson:2011}. \citet{Tsuboi:2016}, \citet{Matsumura:2011}, and  \citet{Kawagoe:2014} detected large flares with luminosities of $2\times10^{32}$~erg~s$^{-1}$ or larger, using MAXI. 

\begin{center}
	\begin{table}[htbp]
	\centering
	\caption{General properties of UX Ari}
	{\begin{tabular}{lcccc} \hline
		Parameters  & Primary & Secondary & References$^*$ \\ \hline
	    Sp. type & K0IV & G5V & (1) \\
	    Radius [\rs] & 5.6~$\pm$~0.1 & 1.6~$\pm$~0.2 & (2) \\
	    Separation [mas] & \multicolumn{2}{c}{1.750~$\pm$~0.01} & (2) \\
	    Separation [\rs] & \multicolumn{2}{c}{18.8~$\pm$~0.1} & (2)(3) \\
	    Distance [pc] & \multicolumn{2}{c}{50.2} & (3) \\
	    Orbital period [d] &
	    \multicolumn{2}{c}{6.43791~$\pm$~0.00006} & (1) \\
	    Inclination [degree] & \multicolumn{2}{c}{125.0~$\pm$~0.5} & (2) \\
		  \hline
	\end{tabular} }
	\label{tb:uxari}
    {$^*$(1)\citet{Carlos:1971}; (2)\citet{Hummel:2017}; (3)\citet{ESA:1997}.}
	\end{table}
	\end{center}

\subsubsection{AR Psc}\label{subsec:ARPsc}

AR Psc (=~HD 8357~=~BD $+$06~211) is a binary system of RS CVn type \citep{Fekel:1996}, located at a distance of 45.9~pc \citep{Gaia:2018}.
It consists of a primary K1 subgiant and a secondary G7V dwarf.
The projected rotational velocity ($v$ $\sin i$) of the AR Psc primary is 6.5~$\pm$~2~km~s$^{-1}$ whereas the orbital period is 14.3023~d \citep{Fekel:1996}. 
AR Psc is an active binary in H$\mathrm{\alpha}$ emission \citep{Fekel:1996}.
\citet{Fekel:1986} found that the
H$\mathrm{\alpha}$ emission is mostly associated with the component originating from the primary star, although
that from the secondary has a weak contribution to it, as seen in
the combined spectrum. 
AR Psc is also known to be active in the X-ray band \citep{Garcia:1980,Kashyap:1999,Shan:2006}. 
\citet{Nakamura:2016} reported on a flare detection by MAXI, revealing a large activity in X-ray flares.

\section{Analysis \& Results} \label{sec:res}

\begin{figure}
    \centering
    \includegraphics[width=8.5cm, angle=0]{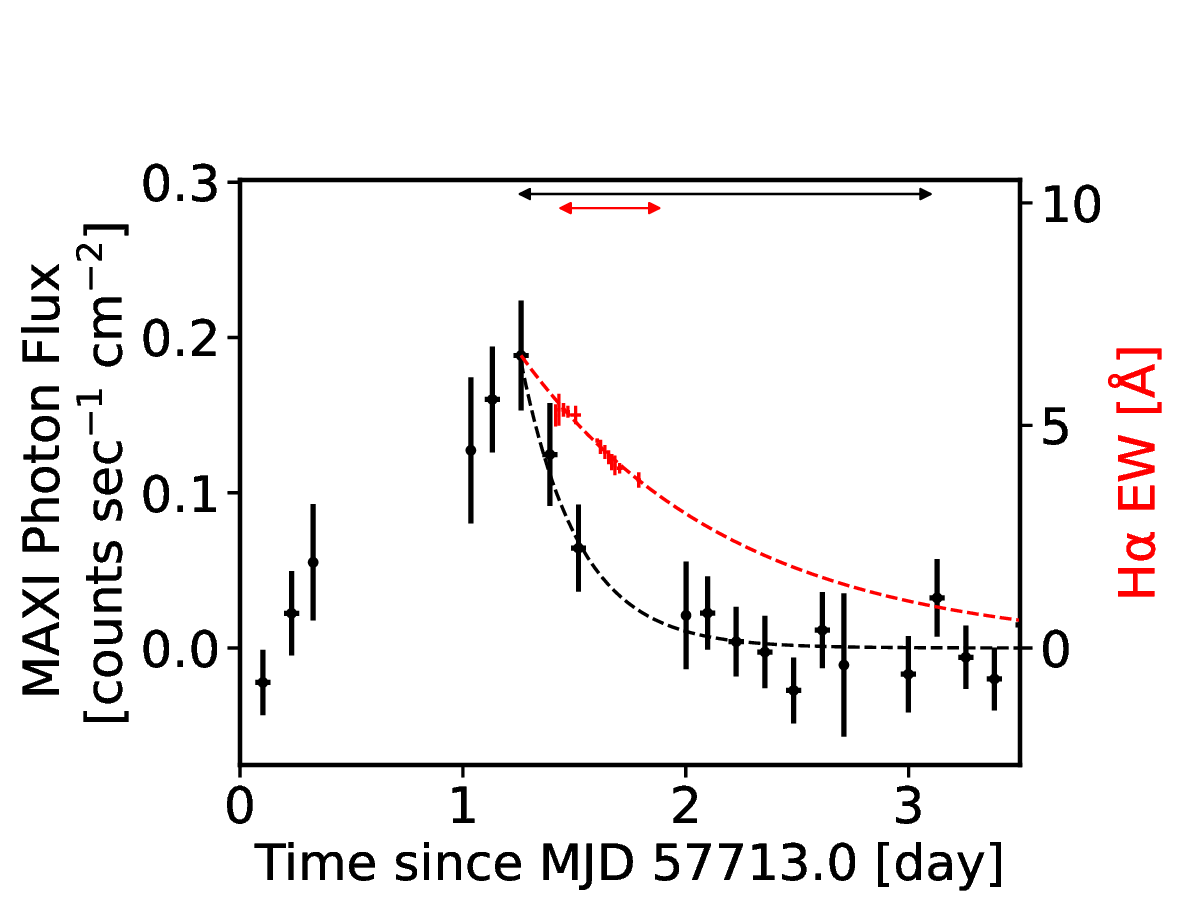}
    \includegraphics[width=8.5cm, angle=0]{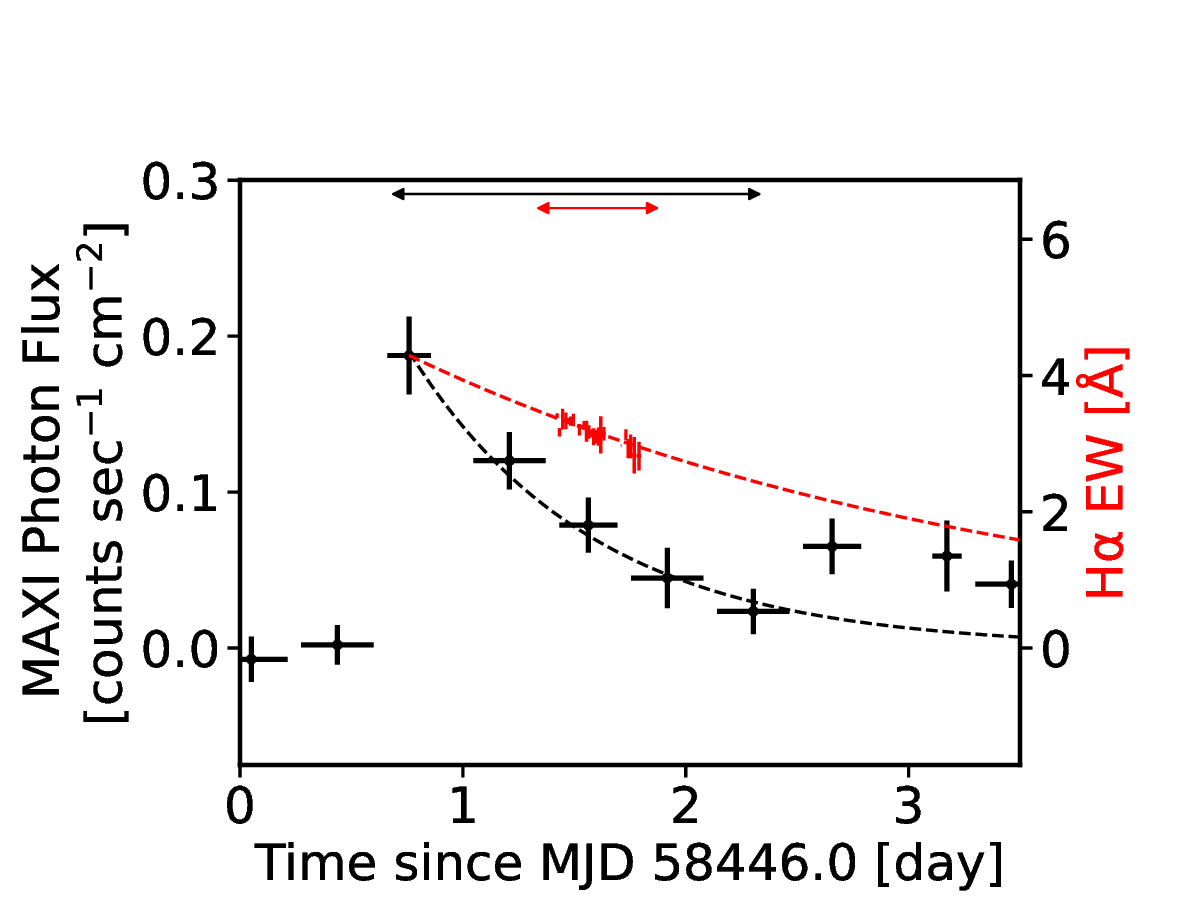}
    \includegraphics[width=8.5cm, angle=0]{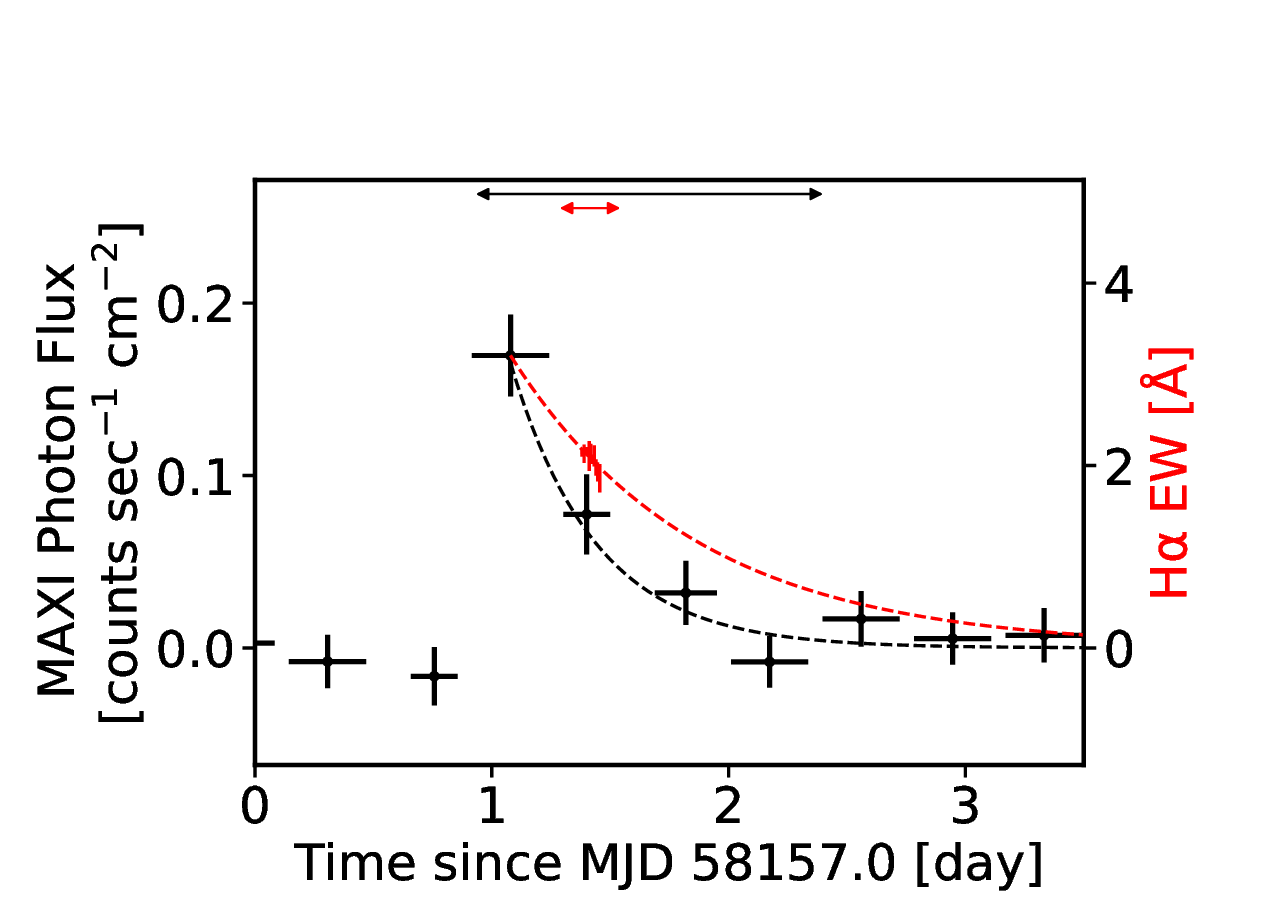}
    \caption{
    SXR and H$\mathrm{\alpha}$ light curves of the Flare 1 (upper panel), Flare 2 (middle), and Flare 3 (lower). The instrumental background (see Section~\ref{subsec:X-ray}) is subtracted in the presented SXR light curves.
  The SXR in the 2--10 keV band, overlaid with its best-fitting model, is plotted in black, with 1~$\sigma$ error, with bins for 2, 6, and 6 MAXI orbital periods in the respective three panels. The time range used for the fitting of each of the SXR and H$\mathrm{\alpha}$ light curves is indicated with horizontal arrows in black and red, respectively, at the top of each panel.
 H$\mathrm{\alpha}$ EW is presented in red with 90\% error bar for the data and best-fitting model.   
    }
    \label{fig:LCs}
\end{figure}

\subsection{Soft X-Ray (SXR) emission}\label{subsec:X-ray}

The MAXI on-demand system \citep{Nakahira:2013} was used for obtaining light curves and spectra. The source photons were extracted from a circular region with a radius of 1$^\circ$.5 centered on each target, i.e., UX Ari and AR Psc, the area of which corresponds to the point-spread function of the detector (Gas Slit Camera) of MAXI. The background photons were extracted from an annular region with inner and outer radii of 1$^\circ$.5 and 4$^\circ$.0, respectively, centered at the source position.

We fitted an exponential model to the light curve of each flare and derived the decay timescale.
The time range used for the fitting of each light curve is shown with a black, horizontal arrow in figure \ref{fig:LCs}.
In the light curves, Flares 2 and 3 may have secondary peak.
Then, we fitted the X-ray light curves, limiting the time region for only $\sim$2 d from the flare peak (see figure~\ref{fig:LCs}) to avoid possible contamination by a potential secondary peak, which may appear then.
Table~\ref{tb:LCfit} lists the fitting results, including the $e$-folding time, $\tau$.  

	\begin{table}[htbp]
	\caption{Flare $e$-folding time $\tau$.$^*$
	}
	\centering
	{\begin{tabular}{lccccc} \hline
		& &Region of&$\tau$&Reduced~$\chi^{2}$ \\
		Flare ID&&interest&& \\
		&&[d] / [$10^{4}$ s]& [$10^{4}$ s]& (d.o.f.) \\
        \hline
	    Flare 1 & SXR & 1.9 / 16 & 2 (1--4) & 0.3 (9) \\
	    & H$\mathrm{\alpha}$ & 0.5 / 4.3 & 8 (7--10) & 0.6 (11) \\
	    Flare 2 & SXR & 1.9 / 16 & 7 (5--10) & 0.2 (3) \\
	     & H$\mathrm{\alpha}$ & 0.4 / 3.5 & 24 (19--32) & 0.8 (22) \\
	    Flare 3 & SXR & 1.5 / 13 & 3 (2--5) & 0.8 (2) \\
	     & H$\mathrm{\alpha}$ & 0.08 / 0.69 & 7 (4--17) & 0.4 (8) \\
		  \hline
	\end{tabular} }
	\label{tb:LCfit}
    {$^*$ 
    In the 2--10 keV band for SXR. 
    The errors are 90\% confidence range.
    ``d.o.f." means the degree of freedom.
        }    
        \end{table}

\begin{figure}
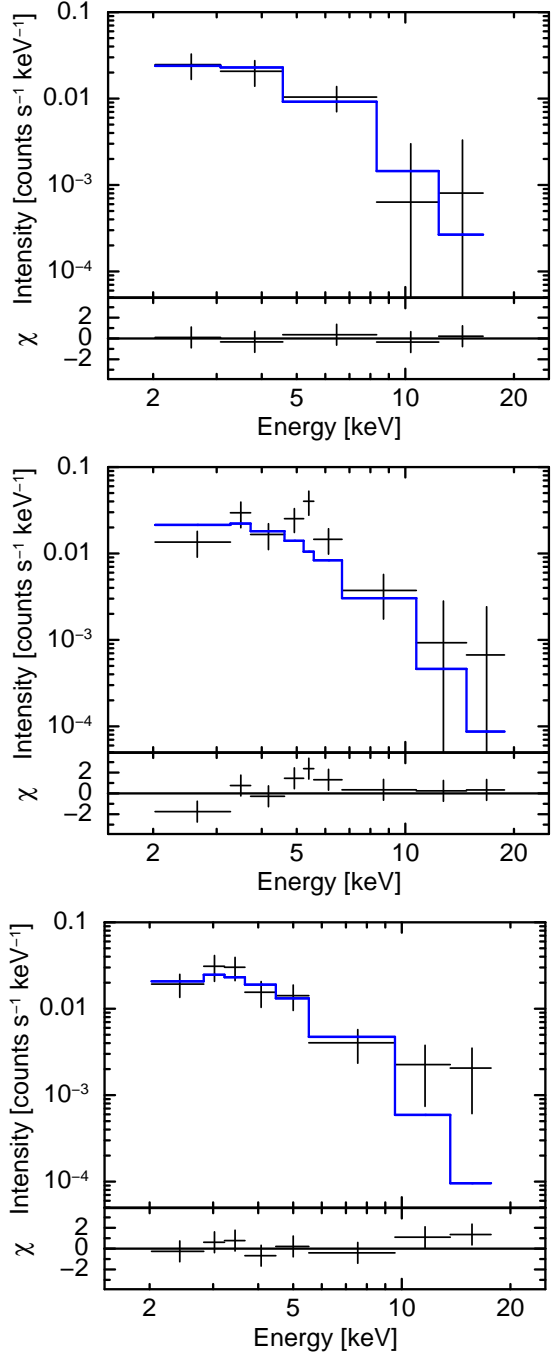

    \centering
    \includegraphics[width=6cm, angle=-90]{Figures/flare1_spec_v13.eps}\\
    \includegraphics[width=6cm, angle=-90]{Figures/flare2_spec_v13.eps}\\
    \includegraphics[width=6cm, angle=-90]{Figures/flare3_spec_v13.eps}
    \vspace*{0.2cm}
    \caption{
    Background-subtracted SXR spectra of Flare 1 (upper panels), Flare 2 (middle), and Flare 3 (lower). The quiescent component is not subtracted. The data points are binned for 2, 6, and 6 MAXI orbits in the respective panels.
    The errors are in 1$\sigma$. The solid lines show the best-fitting models. The residuals for the best-fitting model are also shown at the lower part of each panel.
    }
    \label{fig:MAXIspec}
\end{figure}

The source spectrum during each flare was extracted from the data of the time bin for the peak flux in each flare (figure~\ref{fig:LCs}) and the background component, accumulated from the above-mentioned annular region, is subtracted from it. {The latter represents the sum of the instrumental and sky background components.
Figure~\ref{fig:MAXIspec} shows the resultant flare spectra. 
Although the spectra contain the quiescent component, the contribution is negligible;
\citet{Webb:2020} presented the catalog of active stars, including our targets, UX Ari and AR Pic, which contains the quiescent fluxes of the sources. Their flux levels are $\sim$3\% or lower of any peak fluxes of the flares we reported here.
}

We fitted the spectra with a thin thermal plasma model, APEC \citep{Smith:2001} in the Xspec package, where the redshift and metal abundace were fixed to 0 and 0.3, respectively. 
In the fitting, as for Flares 1 and 2, we fixed the temperature $kT$ to the average value according to \citet{Tsuboi:2016} because otherwise no meaningful constraints would be obtained due to poor data statistics.
Table~\ref{tb:SPECfit} lists the fitting results and figure~\ref{fig:MAXIspec} shows the best-fitting models. 
Then, the radiation energies were calculated, from the best-fitting $e$-folding timescale and the luminosity, and are given in table~\ref{tb:giant_flares}. 

\begin{center}
	\begin{table}[htbp]
	\caption{SXR best-fitting spectral parameters.$^*$}
	\centering
	{\begin{tabular}{lcccc} \hline
		 & kT & Emission measure & Reduced~$\chi^{2}$ \\
	    Flare ID & [keV] & [$10^{55}$~cm$^{-3}$] & (d.o.f.)  \\ \hline
	    Flare 1 & 3.5$^{\dag}$ & 5 (3--6) & 0.1 (4) \\
	    Flare 2 & 3.5$^{\dag}$ & 4 (3--5) & 1.7 (8) \\
	    Flare 3 & 3 (2--6) & 4 (3--7) & 0.8 (6) \\
		  \hline 
	\end{tabular} }

    {$^*$ Abundance and redshift are fixed at 0.3 and 0, respectively. The errors indicate a 90\% confidence range. ``d.o.f". means the degree of freedom. \\
    $^{\dag}$ kT is fixed according to \citet{Tsuboi:2016} (see text for detail).
        }    
	\label{tb:SPECfit}
	\end{table}
\end{center}

\begin{table*}[htbp]
	\centering
	\caption{
	Summary of the three flares.
	}
	\scriptsize
	\renewcommand\arraystretch{2}
	{\fontsize{9pt}{6pt}\selectfont
	\begin{tabular}{lcccccccc} \hline
		&  & \multicolumn{2}{c}{{Flare peak time}} & & Peak luminosity$^{\dag}$ &   Energy$^{\dag}$ \\
		Flare  ID & Star & UT & MJD & Wavelength & [$10^{32}$ erg~s$^{-1}$] & [$10^{36}$ erg]    \\ \hline
		Flare 1 & UX Ari & 2016/11/22 06:16 & 57714.2612 & SXR & 3 (2--4) & 6 (2--12) \\
	    && && H$\mathrm{\alpha}$ & 0.13 (0.06--0.20) & 1.1 (0.5--1.7) \\
	    Flare 2 & UX Ari & 2018/11/24 18:12 & 58446.7587 & SXR & 3 (2--4) & 20 (10--30) \\
	    && && H$\mathrm{\alpha}$ & 0.08 (0.04--0.12) & 1.9 (0.9--3.0) \\
	    Flare 3 & AR Psc & 2018/02/09 01:53 & 58158.0790 & SXR & 2.5 (1.4--3.1) & 8 (3--13) \\
	    && && H$\mathrm{\alpha}$ & 0.04 (0.02--0.07) & 0.3 (0.1--0.8) \\
		  \hline 
    
	\end{tabular}
    {\\
    $^*$ In the 2--10 keV band for SXR. \\
    $^{\dag}$ The errors are 90\% confidence range.
    The H$\mathrm{\alpha}$ luminosities represent the values extrapolated to the times of the X-ray peak times. 
The errors for the H$\mathrm{\alpha}$ luminosities and radiated energies include the uncertainties in determining the decay times.
        }    
	}
	\label{tb:giant_flares}
\end{table*}

\subsection{H$\mathrm{\alpha}$ emission}\label{subsec:Ha}

{
We derived the flux of the H$\mathrm{\alpha}$ emission line for each flare as follows. In general, the direct measurement of the flux of optical grating data is known to be very difficult due to unstable pointing of a telescope and/or weather condition. Hence, we instead derived an equivalent width (EW) of H$\mathrm{\alpha}$ first. Then we converted the H$\mathrm{\alpha}$ EW to its flux. 
}

To obtain the EW of the H$\mathrm{\alpha}$ emission line, we took each frame of the SCAT observations of UX Ari and AR Psc with the exposure of 300~s. Also, we took the frame of 73 Cet (HD~15318) and 2 Ori (HD~30739) in the same way as the standard stars for UX Ari and AR Psc, respectively, for photometric corrections.
We reduced the SCAT data frame-by-frame with a standard manner in Python\textquotesingle s astronomical packages. These reductions are all common among the observations of the flare sources in both flare phases and quiescent phases.

\begin{figure}
    \centering
    \includegraphics[width=8cm, angle=0]{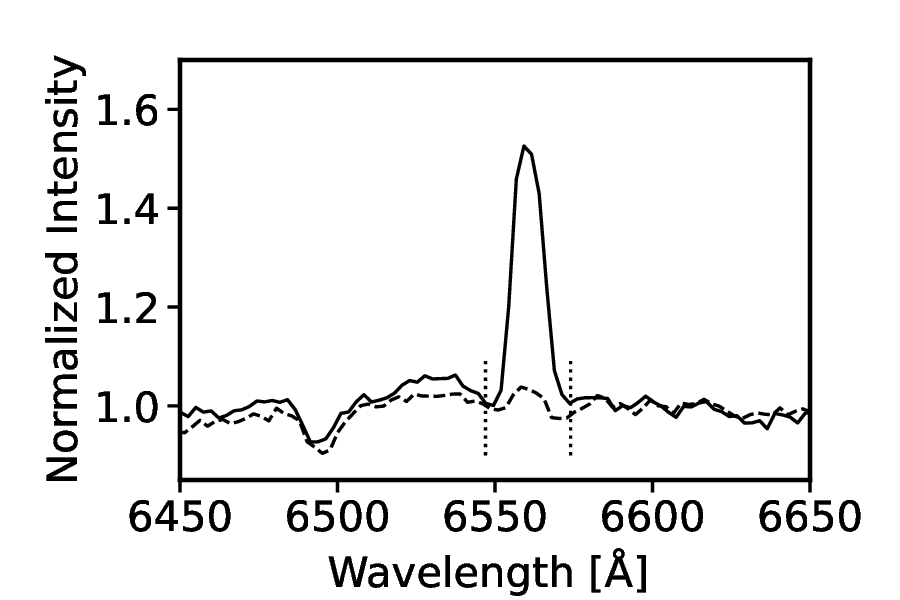}
    \caption{
    Normalized spectra in a wavelength range at around the H$\mathrm{\alpha}$ emission line. The solid and dashed lines correspond to Flare 1 and quiescent components, respectively. See text for details. 
    }
    \label{fig:haspec}
\end{figure}

Figure~\ref{fig:haspec} shows an example of spectra during flare phase and quiescent phase, but normalized at 6,600\AA. 
We applied a standard technique to our spectra and derived the EW. We fitted the H$\mathrm{\alpha}$ emission line with a Gaussian, and picked up the $\pm3\sigma$ points from the Gaussian peak as the ``edge" between the emission line and continuum emission (figure~\ref{fig:haspec}).
As for the H$\mathrm{\alpha}$ line intensity, we integrated the data above the linear function which connects both sides of the ``edge". As for the continuum level, we took the value of the linear function, just at the center of the Gaussian. We then divided the integrated H$\mathrm{\alpha}$ line intensity by the continuum level, and finally obtained the EW. 

Since UX Ari and AR Psc had orbital variation in EW during the quiescent phase, we needed to determine the curve to subtract the orbital variation.
We first folded their light curves, in the quiescent phase, with the rotation period $P_{\mathrm{rot}}$, 
and fitted it with
a function of $a \sin[(x-b)/P_{\mathrm{rot}}]+C$, and determined the EW in each orbital phase.
Table~\ref{tb:Ha_sin} lists the best-fitting values of $a$, $b$, and $C$. 
The value of the EW curve at the same orbital phase was then subtracted from that during the flare and was defined as the EW of the flare. Here, many past optical observations of solar and stellar flares show no large change in continuum flux \citep[e.g.][]{Namekata:2020}. We then assume that the continuum flux of our observations did not change from the quiescent to the flare phase. The fact that no significant variation was found for the continuum flux in the band adjacent to the H$\mathrm{\alpha}$ line supports the assumption. As a result, the H$\mathrm{\alpha}$ EW is basically proportional to the flux of the band where the line component dominates. The resultant curve for the EW is plotted in figure \ref{fig:LCs}.

We fitted the H$\mathrm{\alpha}$ light curves with an exponential function in the same manner for the SXR light curves.  Table~\ref{tb:LCfit} lists the fitting result.
Here, as for the H$\mathrm{\alpha}$ light curve of Flare 1, we performed fitting for the time region of the first day of the flare peak only, during which an enhancement in emission was apparent. We note that the inclusion of the data in the second observation period (on the fourth day) in the fitting yielded the best-fitting decay time of 9.9 (9.5--10.4) $\times 10^4$ s, which is not significantly different from the original result of 8 (7--10) $\times 10^4$ s (table~\ref{tb:LCfit}); in other words, the selection of the time region between the two for the fitting has no significant influence on the result.

In order to obtain the released energy during a flare as the H$\mathrm{\alpha}$ line emission, we needed to estimate the peak H$\mathrm{\alpha}$ intensities, although they are not covered in our follow-up observations. Then, we needed to extrapolate the best-fitting function to the flare-peak epoch obtained in the SXR light curve. 
We note that it should be a reasonable assumption, given that the coincidence of the SXR and H$\mathrm{\alpha}$ peaks have been reported from the data of roughly 200 solar flares \citep{Veronig:2002b} as well as from stellar flares \citep{Doyle:1988B, Fuhrmeister:2011, Namekata:2020}, and also 
that the calculated total radiated energy is not very sensitive to the model shape (e.g., whether it is an $e$-folding decay as we employed or two-peak decay, which may be the case in some flares) that is applied to approximate the light curve.

We calculated the H$\mathrm{\alpha}$ peak luminosity by extrapolating the time-profile of the observed H$\mathrm{\alpha}$ EW to the X-ray peak time, subtracting the averaged continuum level during a one-night observation, and using the distance to the object. 
The absolute flux measurements of continuum emission is in general known to be very difficult for the spectroscopic data that we took. The continuum level of our data was actually fluctuated in 50\% at 90\% error. We then gave this error to the H$\mathrm{\alpha}$ flux of our measurements. 
Table~\ref{tb:giant_flares} shows the results, where the uncertainties in determining the decay times are taken into account in the calculation of the errors.

	\begin{table}[htbp]
	\caption{EW of H$\mathrm{\alpha}$ during the quiescent phase.}
	\centering
	{\begin{tabular}{lccccc} \hline
		Flare ID & Star$^*$ & Observation dates & a$^{\dag}$ & b$^{\dag}$ & C$^{\dag}$ \\ \hline
	    Flare 1 & UX Ari & 2016 Oct -- 2017 Mar & 1.0 & 0.3 & 0.6 \\
	    Flare 2 & UX Ari & 2018 Sep -- 2019 Apr & $-$0.2 & 0.0 & 0.4 \\
	    Flare 3 & AR Psc & 2016 Oct -- 2017 Feb, & 0.4 & 0.2 & 0.1 \\
	    &&2018 Feb&&& \\
		  \hline
	\end{tabular} }
    {
    $^*$UX Ari and AR Psc observations folded with ephemerides MJD = 45695.6142 and 46079.450, respectively \citep{Alekseev:2014, Carlos:1971, Fekel:1996}. 
    $^{\dag}$We fitted the light curve to a function of $a \sin[(x-b)/P_{\mathrm{rot}}]+C$.
    }
	\label{tb:Ha_sin}
	\end{table}

\section{Discussion} \label{sec:dis}

\subsection{Flare loop} \label{sec:loop}

We determined loop lengths of flares, using the "$kT$-$EM$ diagram" model formulated by \citet{Shibata:1999}, where the pre-flare electron density was assumed to be $10^9$ cm$^{-3}$. %The loop lengths are about two times longer than the stellar radius for all the three flares. 
Independently, we also derived the same parameters, using the model formulated by \citet{Haisch:1983}.
The estimated loop lengths based on the former model were roughly three times larger than the binary separation for Flares 1 and 2, whereas those based on the latter model are comparable to the binary separation. Table~\ref{tb:loop} summarizes the result.

	\begin{table*}[htbp]
%	\caption{Estimated loop lengths of three flares}
	\caption{Loop lengths of the three flares.}
	\centering
	{\begin{tabular}{lcccccc} \hline
		%Flare ID & \multicolumn{2}{c}{Loop lengths} \\
		%& [\rs] & (Binary separation) \\
        %\hline
	    %Flare 1 & 62 (51--74) & 3.3 (2.7--3.9) \\
	    %Flare 2 & 57 (50--67) & 3.1 (2.7--3.6) \\
	    %Flare 3 & 80 (30--220) & -- \\
		& \multicolumn{4}{c}{Loop lengths$^*$} \\
		& \multicolumn{2}{c}{\citep{Shibata:1999}} & \multicolumn{2}{c}{\citep{Haisch:1983}} \\
		Flare ID & [\rs] & (Binary separation) & [\rs] & (Binary separation) \\
        \hline
	    Flare 1 & 62 (51--74) & 3.3 (2.7--3.9) & 11 (7--16) & 0.6 (0.4--0.9) \\
	    Flare 2 & 57 (50--67) & 3.1 (2.7--3.6) & 25 (20--32) & 1.3 (1.1--1.7) \\
	    Flare 3 & 80 (30--220) & -- & 14 (10--19) & -- \\
	    \hline
	\end{tabular} }
	    %\begin{tabnote} 
	    \\
    {$^*$The errors indicate the 90\% confidence range.
        }
        %\end{tabnote} 
	\label{tb:loop}
	\end{table*}

\subsection{Radiation energy}\label{subsec:energy}

\citet{Butler:1988} presented the linear relationship between the radiation energy of flares in the H$\mathrm{\gamma}$ emission line ($E_{\mathrm{H\gamma}}$) and that in the X-ray band in the 0.04--2~keV band. \citet{Butler:1993} further extended the relation in energy to the range of seven orders of magnitude: $10^{29}$--$10^{36}$ erg
\footnote{The ``integrated flux'' with the notations $L_\mathrm{X}$ and $L_{\mathrm{H\gamma}}$ in \citet{Butler:1988, Butler:1993} should be interpreted as, though not explicitly explained in their papers, the radiated energy \textit{integrated over the flare duration}. This means that its dimension is energy (erg), as opposed to the nominally expected units of energy per unit time (erg s$^{-1}$) for $L_*$. See the definitions of the parameters in e.g., \citet{Thomas:1971}, from which \citet{Butler:1993} adopted the values, for verification.
%We regard the "integrated flux" Lx and Lgamma in \citep{Butler:1988, Butler:1993} as the energy integrated during the flare duration time, from the points of view of the unit they used for the parameters, erg, and the actual, observed values for the solar flares.
}.

\begin{figure}
    \centering
    \includegraphics[width=8.5cm, angle=270]{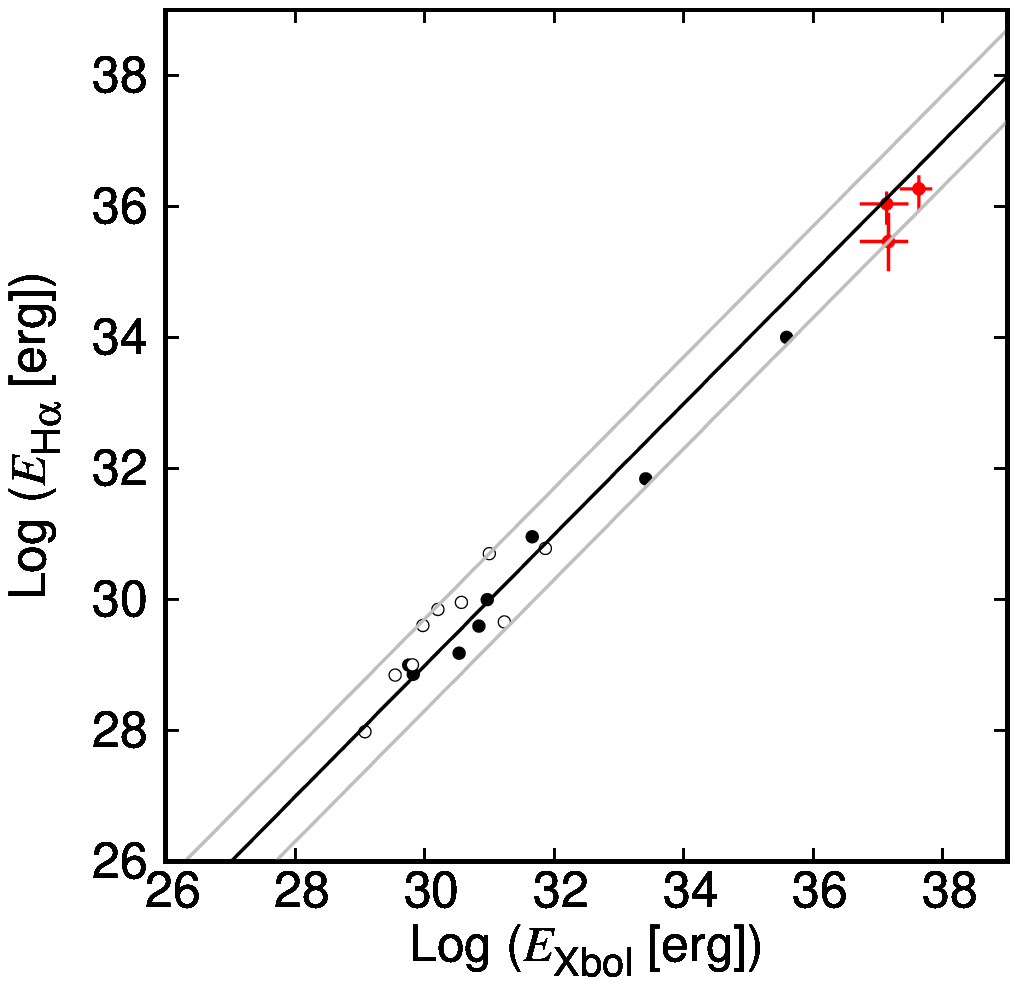}
    \caption{
    Log-log plot of radiation energies of H$\mathrm{\alpha}$ and SXR (0.1--100~keV).  The open and filled circles in black show the data for the solar and stellar flares found in literature. The red data are our results for the MAXI flares with 90\% confidence errors. See text for details. The gray lines indicate 1.6 times of the standard deviation ($\sim$ 90\%) of the data points around the model.
    }
    \label{fig:Ene}
\end{figure}

We produce the same kind of plot, the radiation energy of flares in the H$\mathrm{\alpha}$ emission line ($E_{\mathrm{H\alpha}}$) vs. that in the X-ray bolometric band ($E_{\mathrm{Xbol}}$), in Figure~\ref{fig:Ene}, incorporating
past studies from literature \citep{Butler:1993,Butler:1988,Thomas:1971,Kahler:1982,Butler:1986,Doyle:1988A,Doyle:1988B,Doyle:1991,Johns-Krull:1997}. In deriving the total $E_{\mathrm{H\alpha}}$ energy, some of the past reports used in the compilation presented the radiation energy in the H$\gamma$ emission ($E_{\mathrm{H\gamma}}$) only and no $E_{\mathrm{H\alpha}}$;
we estimated the latter from the former according to the assumed relation $E_{\mathrm{H\alpha}} = 2 \times E_{\mathrm{H\gamma}}$ in such cases, as \citet{Butler:1993} converted the latter to the former using the same conversion factor.
As for the total SXR energy, the flare parameters measured in SXR are converted to those in the bolometric energy band (0.1--100~keV band), using the same method described in \citet{Tsuboi:2016}; the peak bolometric luminosities were obtained, integrating the best-fitting spectral parameters with a thin-thermal plasma model for a range of 0.1--100~keV, and then the bolometric X-ray energies $E_{\mathrm{Xbol}}$ were calculated from it multiplied by its $e$-folding times. Table~\ref{tb:allflares} summarizes the compiled parameters for the stellar flares. %, {the caption of which explains the detailed method of the calculation from the values in each of the references.}
\ref{sec_appendix2} gives a detailed description about the conversion methods for some of the values in table~\ref{tb:allflares} taken from literature.

Fitting all the data points with a linear model with a slope of unity yields the best-fitting relation of

\begin{equation}{
%    E_{\mathrm{Xbol}} = 10 \times E_{\mathrm{H\alpha}},
%     E_{\mathrm{H\alpha}} = 0.1 \times E_{\mathrm{Xbol}}.
    \log{E_{\mathrm{H\alpha}}} = 1 \log{E_{\mathrm{Xbol}}} - 1.0\pm0.7 \ \ ,
    \label{eq:cor_E}}
    \end{equation}
which is shown in figure~\ref{fig:Ene}. The error $\pm0.7$ is calculated as the scatter of the data points around the derived model function in 1.6$\sigma$.  
 
Our result finds that the applicable range of the relation extends for a further two orders in energy, $10^{29}$--$10^{38}$ erg. 
The scattering of the data around the best-fitting linear function is limited within an order. 

\begin{center}
	\begin{table*}[htbp]
	\caption{SXR and H${\alpha}$ radiation energy and $e$-folding time of large flares.$^*$
	}
	%\begin{center}
	{
	\begin{tabular}{lccccc} \hline
	    \centering
		Star & log ($E_{\mathrm{Xbol}}$ [erg] ) & log ($E_{\mathrm{H\alpha}}$ [erg] ) & log ($\tau_{\mathrm{SXR}}$ [s] ) & log ($\tau_{\mathrm{H\alpha}}$ [s] ) & References$^{\dag}$ \\
	    \hline
	    UX Ari & 37.1 (36.7--37.5) & 36.0 (35.7--36.2) & 4.4 (4.1--4.6) & 4.91 (4.86--4.98) & This work (Flare 1) \\
	    & 37.6 (37.3--37.8) & 36.3 (35.9--36.5) & 4.9 (4.7--5.0) & 5.4 (5.3--5.5) & This work (Flare 2) \\
	    AR Psc & 37.2 (36.7--37.5) & 35.5 (35.0--35.9) & 4.5 (4.3--4.7) & 4.8 (4.6--5.2) & This work (Flare 3) \\
	    II Peg & 35.6 & 34.0$^{\ddag}$ & 3.6 & 3.7$^{\ddag}$ & (1), (2) \\
        Gliese 644 & 33.3 & 31.8 & 3.0 & 3.3 & (3) \\
        YZ CMi & 31.0 & 30.0$^{\ddag}$ & 2.4 & 2.0$^{\ddag}$ & (4) \\
        & 31.7 & 31.0$^{\ddag}$ & 2.7 & 2.4$^{\ddag}$ & (5) \\
        UV Cet & 29.8 & 29.0$^{\ddag}$ & 2.0 & 2.4$^{\ddag}$ & (6), (7) \\
        & 30.5 & 29.2$^{\ddag}$ & 2.4 & 2.4$^{\ddag}$ & (6), (7) \\
        & 29.8 & 28.9$^{\ddag}$ & 2.4 & 2.4$^{\ddag}$ & (6), (7) \\
        & 30.8 & 29.6$^{\ddag}$ & 3.0 & 2.7$^{\ddag}$ & (6), (7) \\
        Sun & 29.1 & 28.0 & 3.3 & 2.7 & (8) \\
        & 29.5 & 28.8 & 3.3 & 2.9 & (8) \\
        & 29.8 & 29.0 & 3.3 & 2.7 & (8) \\
        & 30.0 & 29.6 & 3.6 & 3.3 & (8) \\
        & 30.2 & 29.8 & 3.6 & 3.1 & (8) \\
        & 30.6 & 30.0 & 3.6 & 3.1 & (8) \\       
        & 31.0 & 30.7 & 3.9 & 3.7 & (8) \\
        & 31.9 & 30.8 & 3.9 & 3.8 & (8) \\
        & 31.2 & 29.7 & 3.5 & 3.5 & (9) \\
		\hline
	\end{tabular} 
	} 
	%\end{center}
	\label{tb:allflares}
    %\begin{tabnote} 
    \\
        $^*$The errors indicate the 90\%-confidence range. See \ref{sec_appendix2} for the detailed method of the calculation. The energy band of $E_{\mathrm{Xbol}}$ is 0.1--100 keV.  \\ 
	    $^\dag$(1) \cite{Butler:1993}, (2) \cite{Doyle:1991}, (3) \cite{Doyle:1988A}, (4) \cite{Doyle:1988B}, (5) \cite{Kahler:1982}, (6) \cite{Butler:1986}, (7) \cite{Butler:1988}, (8) \cite{Thomas:1971}, (9) \cite{Johns-Krull:1997}\\
	    $^{\ddag}$$E_{\mathrm{H\alpha}}$ $=$ $2E_{H\gamma}$ and $\tau_{\mathrm{H\alpha}}$ $=$ $\tau_{\mathrm{H\gamma}}$ are assumed.  \\
	 \\
    %\end{tabnote} 
	\end{table*}
	\end{center}

\subsection{$e$-folding time}\label{subsec:decay_time}

{
The SXR light curve has limited photon statistics whereas the H$\mathrm{\alpha}$ light curve covers a limited period during each flare (figure~\ref{fig:LCs}). Both limitations contribute to the uncertainty in the $e$-folding times that we derived in section \ref{sec:res}. 
}

{
As for the SXR, the light curves of some large flares have been shown to contain two components with different $e$-folding times \citep[e.g.][]{Agrawal:1988, Tsuboi:1998}. The decay time of the component that is dominant near the peak is reportedly shorter than the one dominant at the tail. 
\citet{Tsuru:1989} reported re-heating or a second flare at the tail of the first and main component in a flare of UX Ari. The decay time that we derived is similar to that of the main flare component ($\sim6.5 \times 10^4$ s) seen in their lightcurve. 
Flare~2 might actually be a double flare similar to the one reported by \citet{Tsuru:1989}, i.e., the SXR light curve of Flare~2 might involve a peak of a possible second flare at around 2 d after the first peak. 
}

{
%Therefore, we conclude that the best-fit parameters for Flare 2 by a single exponential model has an uncertainty of a factor of two. 
}

{
With the optical continuum band, intensive studies of flare decay timescales have been made for solar flares. For example, \citet{Kashapova:2021} reported two decay components, of which the slower-decay component lasts longer by a factor of 2--10 than the faster-decay one. 
Our data of H$\mathrm{\alpha}$ covered a small fraction of the flare decay phase, suggesting that we only detected one component out of possibly two or more components in reality in the light curves.  
Therefore, we conclude that the best-fitting $e$-folding times of the SXR and H$\mathrm{\alpha}$ emissions that we measured with the single-exponential model fitting has a (systematic) error of a factor of 2--10 and thus is robust in only one order of accuracy. 
}

{
Many studies of stellar flares \citep[e.g.,][]{Kashapova:2021} show that faster and slower decaying components have high and low peak fluxes, respectively. Then, the flare energy that is determined from the decay time multiplied by the peak luminosity is not varied much because the uncertainties compensate each other, if to a certain degree.
%The systematic uncertainty of the energy $E$ is expected as smaller than that of the $e$-folding time $\tau$. 
}

\begin{figure}
    \centering
    \includegraphics[width=8.5cm, angle=270]{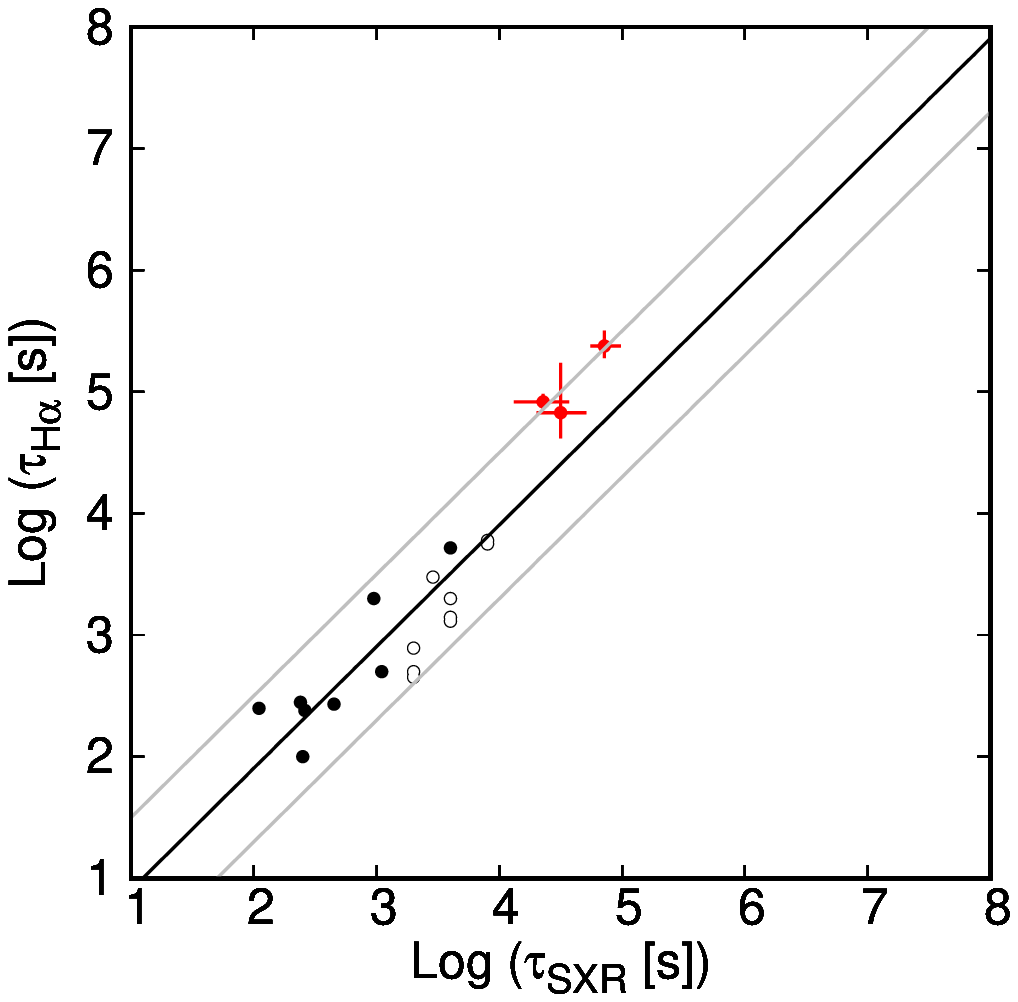}
    \caption{Log-log plot of $e$-folding times of H$\mathrm{\alpha}$ and SXR. 
    %H$\mathrm{\alpha}$線とX線の減衰時間のLog-log plot。
    The symbols are the same as in Figure~\ref{fig:Ene}. The gray lines indicate 1.6 times of the standard deviation ($\sim$ 90\%) of the data points around the model. }
    \label{fig:Tau}
\end{figure}

\citet{Veronig:2002} found a linear correlation between the $e$-folding times $\tau_{\mathrm{SXR}}$ and $\tau_{\mathrm{H\alpha}}$ for solar flares.  
In Figure~\ref{fig:Tau}, we compiled the same kind of plot, but incorporating the data for stellar flares as in the previous subsection. Here, the reported $e$-folding times were adopted as they are, regardless of the energy band used in each study.
The linear relationship is again apparent in this figure.
We then fitted the plot with a linear function.  The best-fit model, which is overlaid in figure~\ref{fig:Tau}, is given by
 \begin{equation}
%%    \tau_{\mathrm{SXR}} = 1.3 \times \tau_{\mathrm{H\alpha}},
%    \tau_{\mathrm{H\alpha}} = 0.8 \times \tau_{\mathrm{SXR}},
    \log{\tau_{\mathrm{H\alpha}}} = 1 \log{\tau_{\mathrm{SXR}}} - 0.1 \pm 0.6 {\ \ },
    \label{eq:cor_T}
    \end{equation}
%が得られた。
where the error is calculated as the scatter of the data points around the derived model function in 1.6$\sigma$.

We find that
all the data fall within an factor of 4 around the best-fitting linear function. 
This result implies that the two parameters $\tau_{\mathrm{SXR}}$ and $\tau_{\mathrm{H\alpha}}$ are almost the same among not only the solar flares but also much more energetic stellar flares. The result extends the applicable range for the relationship by
an order of magnitude from the previous result \citep{Veronig:2002}. 
It is notable that the overall trend of the data points in the $\tau_\mathrm{SXR}$-$\tau_\mathrm{H\alpha}$ space broadly follows a linear relation for 3 orders of magnitude for the duration $\tau$.

\subsection{Solar to large stellar flares}

We have found proportional relationships between the SXR and H$\mathrm{\alpha}$ emission line for the flare radiation energy $E$ and $e$-folding time $\tau$ from the solar to large stellar flares. 
%We here compare the large stellar flares with the solar flares. 
As for the Sun, thanks to the close distance to it, many images have been obtained with high spatial resolution in the H$\mathrm{\alpha}$ emission line and SXR band \citep[e.g.][]{Asai:2004, Jing:2016}.
SXRs are dominated by optically-thin thermal emission from high-temperature plasma of tens of millions of degrees that fills the loop of each flare \citep[e.g.][]{Shibata:2011}. 
The H$\mathrm{\alpha}$ line is, by contrast, found to be emitted near the foot-points of a flare loop and/or from the inner side of SXR flare loop, called postflare loop \citep[see figure 42 in ][]{Shibata:2011}. 
The hydrogen responsible for the H$\mathrm{\alpha}$ emission line is thought to be excited by one or all of (1) the collision of accelerated particles, (2) heat conducted from the high-temperature plasma at the foot-point of the flare loop, and (3) the high-energy radiation from the high-temperature plasma.  
The nominal loop size of the solar flare is reported to be $\sim$ 0.1 {\rs} \citep[e.g.][]{Kontar:2011}. Although the loop sizes derived for Flares 1 and 2 are both estimated to be more than twice the stellar radius, our finding of the scaling relationships suggests that the origins of the SXR and H$\mathrm{\alpha}$ of the large flare are common to those of the Sun.

In this work, the studied correlation is limited to $\mathrm{H\alpha}$ and X-rays. Although some studies have been done about the correlation between white light and X-rays, the uncertainty in the results is considerable mainly due to their limited radiated-energy ranges of the sample stellar flares \citep[e.g.,][]{Flaccomio:2018, Guarcello:2019, Kuznetsov:2021}. %{\citep[e.g.,][]{2018A&A...620A..55F}}. 
Future simultaneous observations of stellar flares with X-rays and other bands, including white light, would provide us with insights about the multi-wavelength nature of stellar flares for a wide energy band.

\section{Summary} \label{sec:sum}

\begin{enumerate}

\item
During the five years from 2016 October to 2021 September, three large stellar flares from UX Ari and AR Psc were simultaneously observed in H$\mathrm{\alpha}$ and SXRs.

\item
The radiation energies of the observed flares were $ 10^{36}$--$10^{37} $~erg in the X-ray band and $ 10^{35}$--$10^{36} $~erg in the H$\mathrm{\alpha}$ emission line. By combining the obtained physical parameters and those in literature for solar and stellar flares, a good proportional relation was obtained between the emitted energies of X-ray and H$\mathrm{\alpha}$ emissions for a flare energy range of $10^{29}$--$10^{38}$ erg. This is the first confirmation of the relationship based on simultaneous X-ray and H$\mathrm{\alpha}$ observations of massive stellar flares. The ratio of the H$\mathrm{\alpha}$ line emission to that of SXR is $\sim$0.1, if we take the energy band from 0.1 to 100 keV to obtain the X-ray luminosity.

%The timescale and radiation energy of SXRs and H$\mathrm{\alpha}$ emission line were simultaneously obtained for the first time for large stellar flares with energies of $\gtrsim 10^{35}$ ergs.

\item
The $e$-folding times in the decay phase of both SXR and H$\mathrm{\alpha}$ flares were obtained to be $10^{4}$--$10^{5}$~s.
By combining the obtained physical parameters and those in literature for solar and stellar flares, it is confirmed that the $e$-folding times in both bands coincide with the range of $1$--$10^4$~sec.

\item
Even very large stellar flares with energies of 6 orders of magnitude larger than the most energetic solar flares follow the same scaling relationships as %established for 
solar and much less energetic stellar flares. This fact suggests that their physical parameters can be estimated on the basis of the known physics of solar and stellar flares.
%%%% YM addded 2022.1.15, edited by MS
{
We stress that this conclusion is derived using the H$\mathrm{\alpha}$ emission data with a limited coverage for flare durations. 
Future simultaneous observations of SXR and H$\mathrm{\alpha}$ emissions from stellar flares with more complete coverage for the flare duration and with higher statistics 
will verify the scaling relationships.
}
%%%%

\end{enumerate}

\section*{Acknowledgments}
We are grateful to an anonymous referee for constructive comments.
We thank R. Iizuka and Y. Sugawara for acquisition, reduction and interpretation of the SCAT data. This research has made use of the MAXI data, provided by the RIKEN, JAXA, and MAXI teams. This research was partially supported by the Ministry of Education, Culture, Sports, Science and Technology (MEXT), Grant-in-Aid Nos.16K17673, 16K17717, 17K05392, 19K21886, 20H04743, 20K20935, and 21H01121.  Y.T. acknowledges financial support by a Chuo University Grant for Special Research. We thank M. Sakano (Wise Babel Ltd.) for the manuscript corrections including English.
%\begin{appendix}

\appendix
%\chapter{Light curves of the four extra flares}
\section{Light curves of the four extra flares}
\label{sec_appendix}

\renewcommand{\thefigure}{A\arabic{figure}}
\setcounter{figure}{0}

\begin{figure*}
  \begin{minipage}[b]{0.45\linewidth}
    \centering
    \includegraphics[width=7.5cm, angle=0]{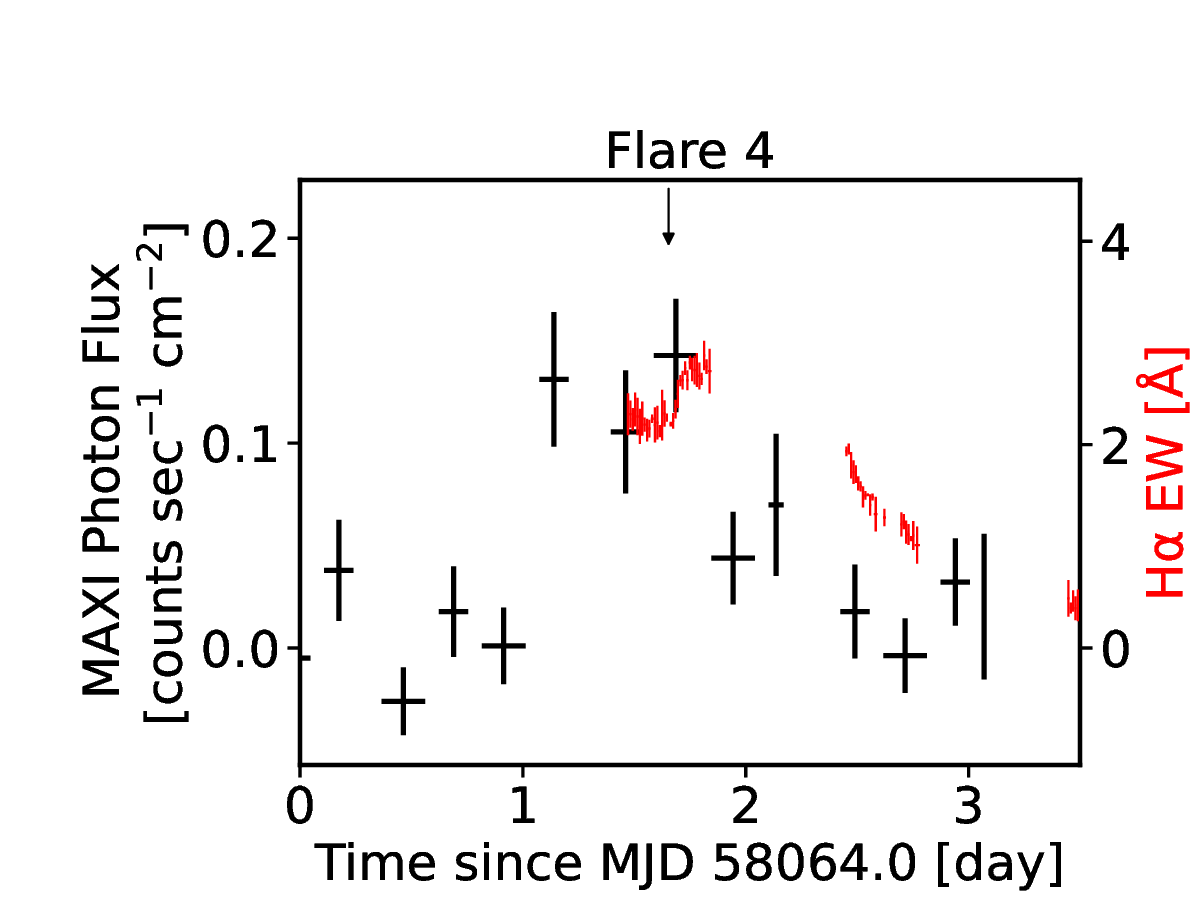}
    \includegraphics[width=7.5cm, angle=0]{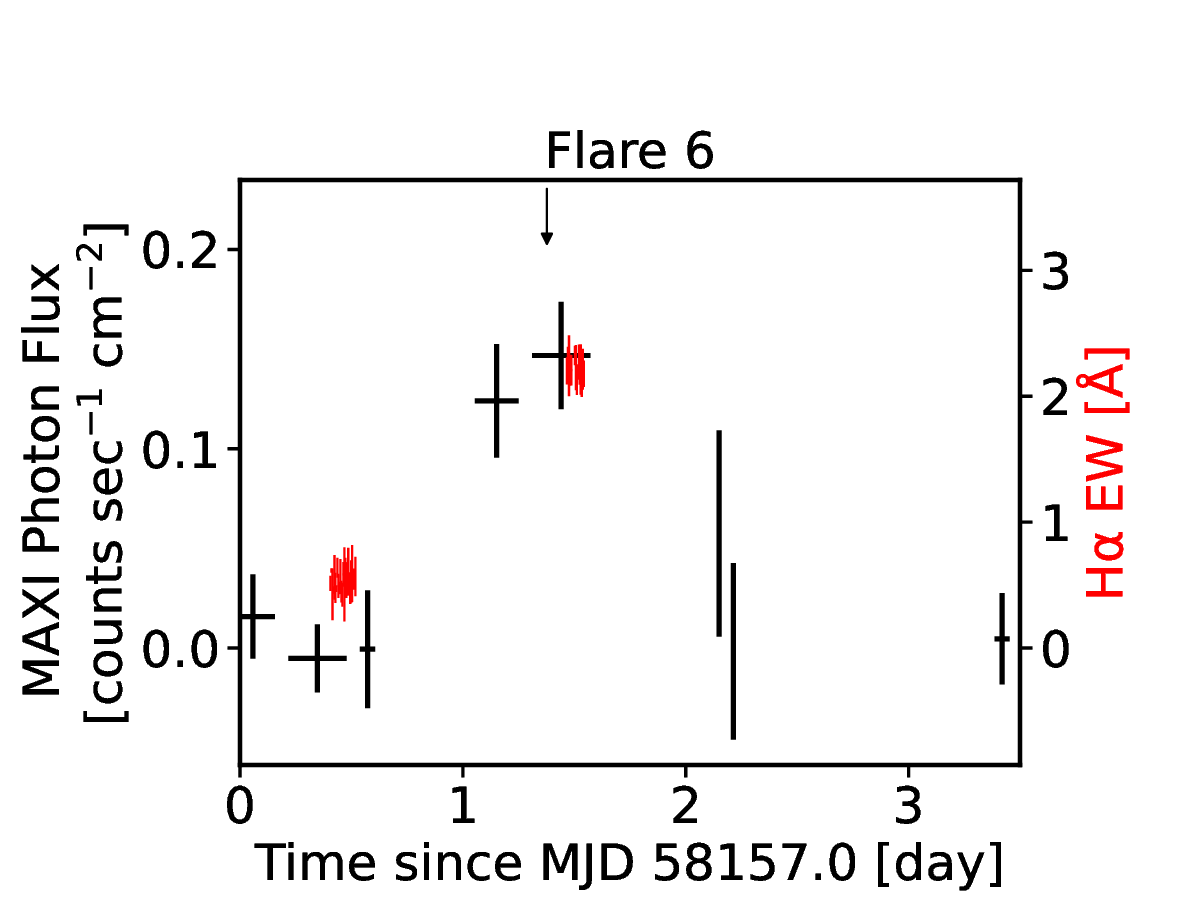}
  \end{minipage}
  \begin{minipage}[b]{0.45\linewidth}
    \centering
    \includegraphics[width=7.5cm, angle=0]{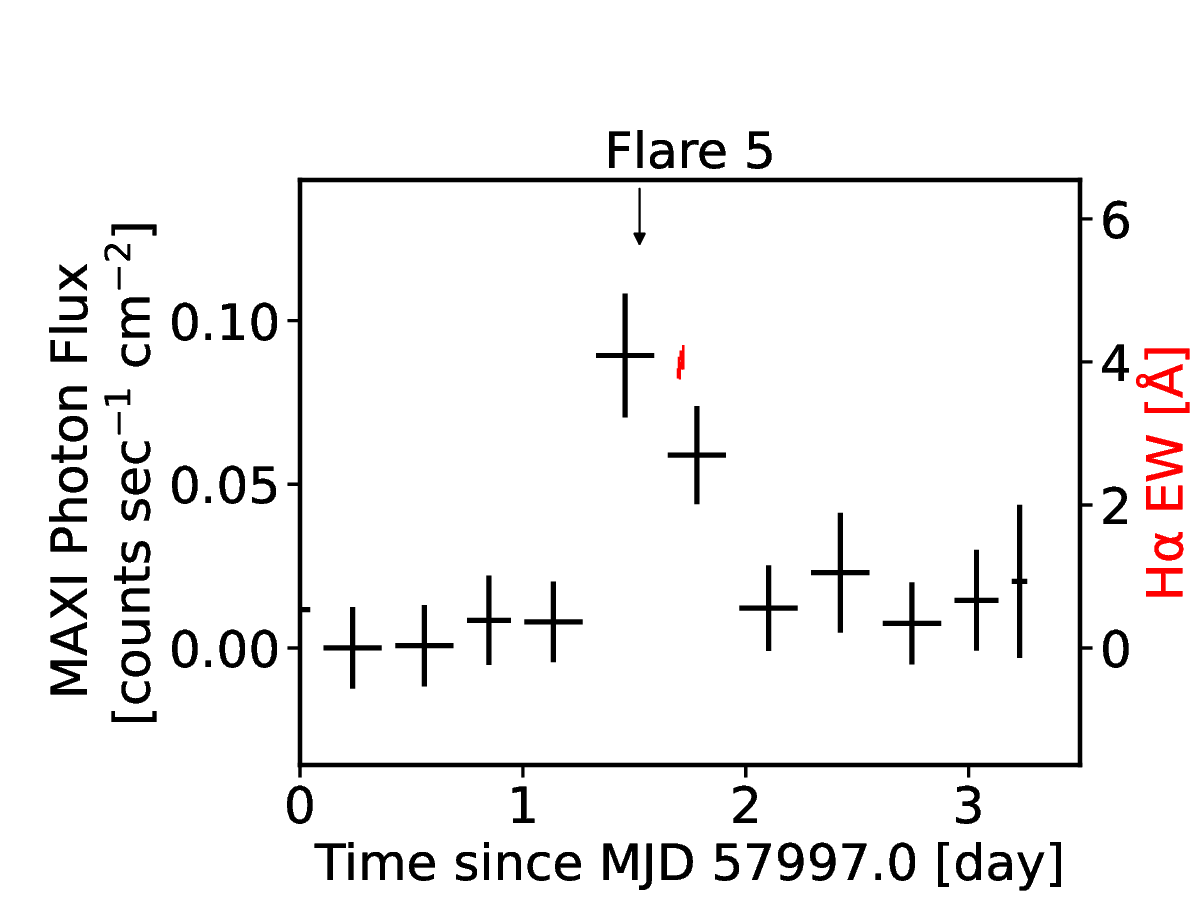}
    \includegraphics[width=7.5cm, angle=0]{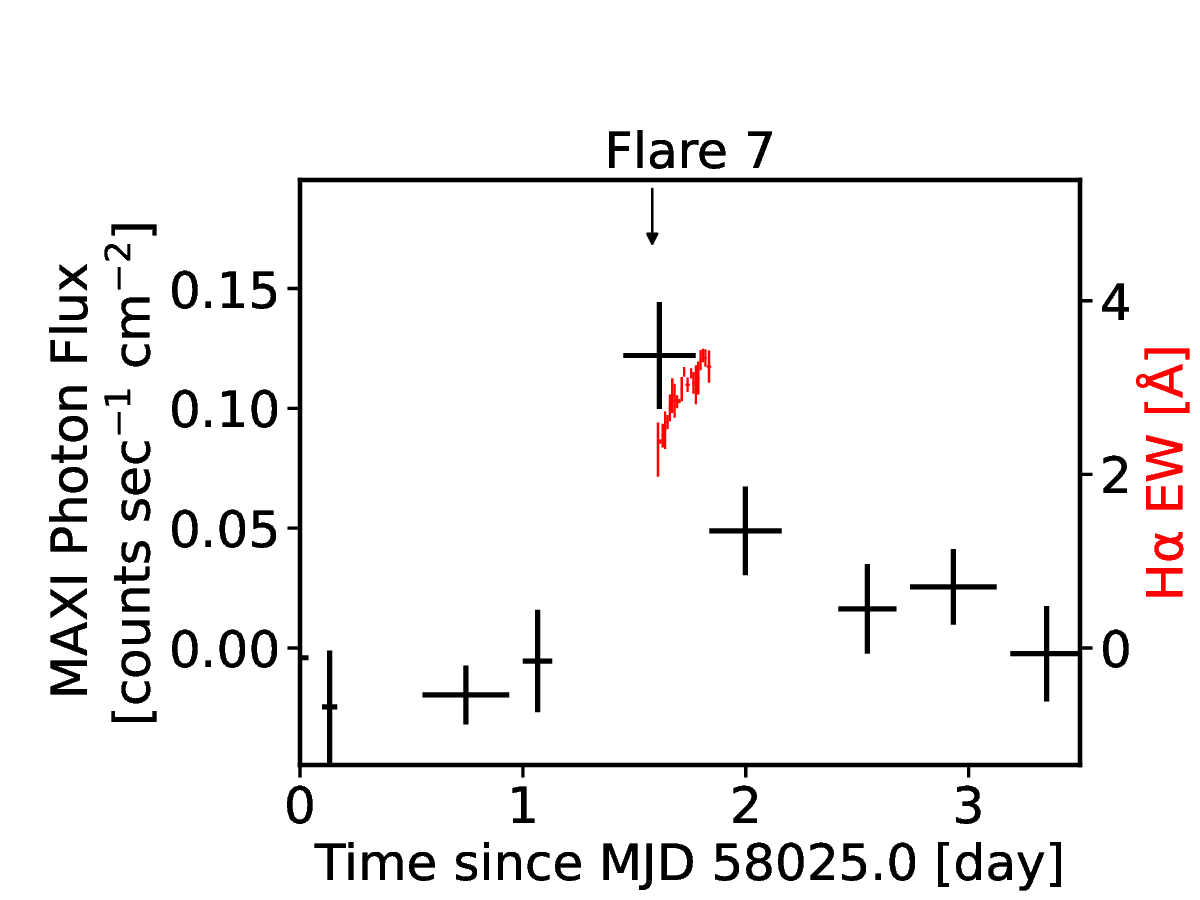}
  \end{minipage}
    \caption{
    SXR and H$\mathrm{\alpha}$ light curves of four of the seven flares that are not discussed in the main text: Flares 4 (upper left), 5 (upper right), 6 (lower left), and 7 (lower right).
    The downward arrow indicates the time of flare detection by ``nova search'' for each flare.
    }
    \label{fig:4LCs}
\end{figure*}

We show in Figure~\ref{fig:4LCs} the light curves of four flares that are not included in our selection for detailed analysis, out of of the seven flares mentioned in the observations. Flare 4 occurred in UX Ari, Flares 5 and 6, in HR1099, and Flare 7, in VY Ari. All the objects are stars classified as RS CVn type.
%\end{appendix}

%\chapter{Calculation of radiation energies and $e$-folding times from those in literature}
\section{Calculation of radiation energies and $e$-folding times from those in literature}
\label{sec_appendix2}

The general procedure to calculate the radiation energies and $e$-folding times in SXRs and H$\mathrm{\alpha}$ (tabulated in table~\ref{tb:allflares}) is described in subsections~\ref{subsec:energy} and \ref{subsec:decay_time}. In some cases, the values are taken from the cited references as they are presented. However, in many cases, some conversions are required to derive the compiled values (table~\ref{tb:allflares}). Here is the detail of the conversion methods. II Peg: $E_\mathrm{Xbol}$ is calculated from the peak X-ray flux (referred to as $F_\mathrm{X,peak}$) and $\tau_\mathrm{SXR}$, both of which are read directly from a figure in the reference, and $\tau_{H\mathrm{\alpha}}$ is read directly from a figure. Glise 644: $E_\mathrm{Xbol}$ is calculated in a similar manner as that for II Peg except that $F_\mathrm{X,peak}$ is taken from a table. YZ CMi with reference (4) in table~\ref{tb:allflares}: $\tau_\mathrm{SXR}$ is calculated from a combination of $F_\mathrm{X,peak}$ directly read from a figure and energy in reference (7), and $\tau_{H\mathrm{\alpha}}$ is from the peak H$\mathrm{\alpha}$ flux ($F_\mathrm{H\alpha,peak}$) from a table and $E_\mathrm{H\alpha}$ in reference (1). YZ CMi with reference (5): $\tau_\mathrm{SXR}$ is from $\tau_\mathrm{SXR}$ calculated from $F_\mathrm{X,peak}$ and $E_\mathrm{X}$ in reference (5), and $\tau_{H\mathrm{\alpha}}$ is directly from a figure. UV Cet: $\tau_\mathrm{SXR}$ is from $F_\mathrm{X,peak}$ and $\tau_\mathrm{SXR}$ is directly from a figure and $\tau_{H\mathrm{\alpha}}$ from a combination of $F_\mathrm{H\alpha,peak}$ directly from a figure and $E_{H\mathrm{\alpha}}$ from reference (1). Sun with reference (8): $E_\mathrm{Xbol}$ is from $F_\mathrm{X,peak}$ directly from a figure, and $\tau_{H\mathrm{\alpha}}$ is found in the same manner as that for UV Cet. Sun with reference (9): $E_\mathrm{Xbol}$ is from a combination of $F_\mathrm{X,peak}$ directly from a figure and $\tau_\mathrm{SXR}$ from $F_\mathrm{X,peak}$ and energy from a table.

\bibliographystyle{elsarticle-harv} 
%% \bibliography{example_updating}
\bibliography{ref}

\begin{thebibliography}{54}
\expandafter\ifx\csname natexlab\endcsname\relax\def\natexlab#1{#1}\fi
\providecommand{\url}[1]{\texttt{#1}}
\providecommand{\href}[2]{#2}
\providecommand{\path}[1]{#1}
\providecommand{\DOIprefix}{doi:}
\providecommand{\ArXivprefix}{arXiv:}
\providecommand{\URLprefix}{URL: }
\providecommand{\Pubmedprefix}{pmid:}
\providecommand{\doi}[1]{\href{http://dx.doi.org/#1}{\path{#1}}}
\providecommand{\Pubmed}[1]{\href{pmid:#1}{\path{#1}}}
\providecommand{\bibinfo}[2]{#2}
\ifx\xfnm\relax \def\xfnm[#1]{\unskip,\space#1}\fi
%Type = Article
\bibitem[{{Agrawal} and {Vaidya}(1988)}]{Agrawal:1988}
\bibinfo{author}{{Agrawal}, P.C.}, \bibinfo{author}{{Vaidya}, J.},
  \bibinfo{year}{1988}.
\newblock \bibinfo{title}{{Binary phase correlated X-ray intensity variations
  and flaring in theRS CVn binary HR 1099.}}
\newblock \bibinfo{journal}{MNRAS} \bibinfo{volume}{235}, \bibinfo{pages}{239}.
%Type = Article
\bibitem[{{Alekseev}(2014)}]{Alekseev:2014}
\bibinfo{author}{{Alekseev}, I.Y.}, \bibinfo{year}{2014}.
\newblock \bibinfo{title}{{Three-Component Model of Spottedness in the
  Classical RS CVn System UX Ari}}.
\newblock \bibinfo{journal}{Ap} \bibinfo{volume}{57}, \bibinfo{pages}{344}.
%Type = Article
\bibitem[{{Asai} et~al.(2004){Asai}, {Yokoyama}, {Shimojo}, {Masuda},
  {Kurokawa} and {Shibata}}]{Asai:2004}
\bibinfo{author}{{Asai}, A.}, \bibinfo{author}{{Yokoyama}, T.},
  \bibinfo{author}{{Shimojo}, M.}, \bibinfo{author}{{Masuda}, S.},
  \bibinfo{author}{{Kurokawa}, H.}, \bibinfo{author}{{Shibata}, K.},
  \bibinfo{year}{2004}.
\newblock \bibinfo{title}{{Flare Ribbon Expansion and Energy Release Rate}}.
\newblock \bibinfo{journal}{ApJ} \bibinfo{volume}{611}, \bibinfo{pages}{557}.
%Type = Article
\bibitem[{{Butler}(1993)}]{Butler:1993}
\bibinfo{author}{{Butler}, C.J.}, \bibinfo{year}{1993}.
\newblock \bibinfo{title}{{An extended correlation between the Balmer and soft
  X-ray emission from solar and stellar flares.}}
\newblock \bibinfo{journal}{A\&A} \bibinfo{volume}{272}, \bibinfo{pages}{507}.
%Type = Article
\bibitem[{{Butler} et~al.(1988){Butler}, {Rodono} and {Foing}}]{Butler:1988}
\bibinfo{author}{{Butler}, C.J.}, \bibinfo{author}{{Rodono}, M.},
  \bibinfo{author}{{Foing}, B.H.}, \bibinfo{year}{1988}.
\newblock \bibinfo{title}{{A correlation between Balmer and soft X-ray emission
  from stellar \& solar flares.}}
\newblock \bibinfo{journal}{A\&A} \bibinfo{volume}{206}, \bibinfo{pages}{L1}.
%Type = Article
\bibitem[{{Butler} et~al.(1986){Butler}, {Rodono}, {Foing} and
  {Haisch}}]{Butler:1986}
\bibinfo{author}{{Butler}, C.J.}, \bibinfo{author}{{Rodono}, M.},
  \bibinfo{author}{{Foing}, B.H.}, \bibinfo{author}{{Haisch}, B.M.},
  \bibinfo{year}{1986}.
\newblock \bibinfo{title}{{Coordinated Exosat and spectroscopic observations of
  flare stars and coronal heating}}.
\newblock \bibinfo{journal}{Nature} \bibinfo{volume}{321},
  \bibinfo{pages}{679}.
%Type = Article
\bibitem[{{Carlos} and {Popper}(1971)}]{Carlos:1971}
\bibinfo{author}{{Carlos}, R.C.}, \bibinfo{author}{{Popper}, D.M.},
  \bibinfo{year}{1971}.
\newblock \bibinfo{title}{{HD 21242, A Spectroscopic Binary with H and K
  Emission}}.
\newblock \bibinfo{journal}{PASP} \bibinfo{volume}{83}, \bibinfo{pages}{504}.
%Type = Article
\bibitem[{{Catalano} et~al.(2003){Catalano}, {Umana}, {Cafra}, {Frasca},
  {Trigilio} and {Marilli}}]{Catalano:2003}
\bibinfo{author}{{Catalano}, S.}, \bibinfo{author}{{Umana}, G.},
  \bibinfo{author}{{Cafra}, B.}, \bibinfo{author}{{Frasca}, A.},
  \bibinfo{author}{{Trigilio}, C.}, \bibinfo{author}{{Marilli}, E.},
  \bibinfo{year}{2003}.
\newblock \bibinfo{title}{{A Simultaneous H$\alpha$ and Radio Flare on the RS
  CVn System UX Ari}}.
\newblock \bibinfo{journal}{in Cambridge Workshop on Cool Stars, Stellar
  Systems, and the Sun, 12, The Future of Cool-Star Astrophysics, eds} ,
  \bibinfo{pages}{981}.
%Type = Article
\bibitem[{{Doyle} et~al.(1988a){Doyle}, {Butler}, {Bryne} and {van den
  Oord}}]{Doyle:1988B}
\bibinfo{author}{{Doyle}, J.G.}, \bibinfo{author}{{Butler}, C.J.},
  \bibinfo{author}{{Bryne}, P.B.}, \bibinfo{author}{{van den Oord}, G.H.J.},
  \bibinfo{year}{1988}a.
\newblock \bibinfo{title}{{Rotational modulation and flares on RS CVn and BY
  DRA systems. VII. simultaneous X-ray, radio and optical data for the dMe star
  YZ CMi on 4/5 March 1985.}}
\newblock \bibinfo{journal}{A\&A} \bibinfo{volume}{193}, \bibinfo{pages}{229}.
%Type = Article
\bibitem[{{Doyle} et~al.(1988b){Doyle}, {Butler}, {Callanan}, {Tagliaferri},
  {de La Reza}, {White}, {Torres} and {Quast}}]{Doyle:1988A}
\bibinfo{author}{{Doyle}, J.G.}, \bibinfo{author}{{Butler}, C.J.},
  \bibinfo{author}{{Callanan}, P.J.}, \bibinfo{author}{{Tagliaferri}, G.},
  \bibinfo{author}{{de La Reza}, R.}, \bibinfo{author}{{White}, N.E.},
  \bibinfo{author}{{Torres}, C.A.}, \bibinfo{author}{{Quast}, G.},
  \bibinfo{year}{1988}b.
\newblock \bibinfo{title}{{Rotational modulation and flares on RS CVn and BY
  DRA systems. VIII. Simultaneous EXOSAT and H alpha observations of a flare on
  the dMe star GL 644 AB (Wolf 630) on 24/25 August 1985.}}
\newblock \bibinfo{journal}{A\&A} \bibinfo{volume}{191}, \bibinfo{pages}{79}.
%Type = Article
\bibitem[{{Doyle} et~al.(1991){Doyle}, {Kellett}, {Byrne}, {Avgoloupis},
  {Mavridis}, {Seiradakis}, {Bromage}, {Tsuru}, {Makishima}, {Makishima} and
  {McHardy}}]{Doyle:1991}
\bibinfo{author}{{Doyle}, J.G.}, \bibinfo{author}{{Kellett}, B.J.},
  \bibinfo{author}{{Byrne}, P.B.}, \bibinfo{author}{{Avgoloupis}, S.},
  \bibinfo{author}{{Mavridis}, L.N.}, \bibinfo{author}{{Seiradakis}, J.H.},
  \bibinfo{author}{{Bromage}, G.E.}, \bibinfo{author}{{Tsuru}, T.},
  \bibinfo{author}{{Makishima}, K.}, \bibinfo{author}{{Makishima}, K.},
  \bibinfo{author}{{McHardy}, I.M.}, \bibinfo{year}{1991}.
\newblock \bibinfo{title}{{Simultaneous detection of a large flare in the X-ray
  and optical regions on the RS CVn-type star II Peg.}}
\newblock \bibinfo{journal}{MNRAS} \bibinfo{volume}{248}, \bibinfo{pages}{503}.
%Type = Article
\bibitem[{ESA(1997)}]{ESA:1997}
\bibinfo{author}{ESA}, \bibinfo{year}{1997}.
\newblock \bibinfo{title}{{The HIPPARCOS and TYCHO catalogues. Astrometric and
  photometric star catalogues derived from the ESA HIPPARCOS Space Astrometry
  Mission}}.
\newblock \bibinfo{journal}{ESA Special Publication, 1200}
  \bibinfo{volume}{1200}.
%Type = Article
\bibitem[{Fekel(1996)}]{Fekel:1996}
\bibinfo{author}{Fekel, F.C.}, \bibinfo{year}{1996}.
\newblock \bibinfo{title}{{Chromospherically Active Stars. XV. HD 8357=AR
  Piscium, an Extremely Active RS CVn System}}.
\newblock \bibinfo{journal}{AJ} \bibinfo{volume}{112}, \bibinfo{pages}{269}.
%Type = Article
\bibitem[{{Fekel} et~al.(1986){Fekel}, {Moffett} and {Henry}}]{Fekel:1986}
\bibinfo{author}{{Fekel}, F.C.}, \bibinfo{author}{{Moffett}, T.J.},
  \bibinfo{author}{{Henry}, G.W.}, \bibinfo{year}{1986}.
\newblock \bibinfo{title}{{A Survey of Chromospherically Active Stars}}.
\newblock \bibinfo{journal}{ApJS} \bibinfo{volume}{60}, \bibinfo{pages}{551}.
%Type = Article
\bibitem[{{Feldman}(1978)}]{Feldman:1978}
\bibinfo{author}{{Feldman}, P.A.}, \bibinfo{year}{1978}.
\newblock \bibinfo{title}{{Large Radio Flares in RS CVn Binaries}}.
\newblock \bibinfo{journal}{BAAS} \bibinfo{volume}{10}, \bibinfo{pages}{418}.
%Type = Article
\bibitem[{{Flaccomio} et~al.(2018){Flaccomio}, {Micela}, {Sciortino}, {Cody},
  {Guarcello}, {Morales-Calderòn}, {Rebull} and {Stauffer}}]{Flaccomio:2018}
\bibinfo{author}{{Flaccomio}, E.}, \bibinfo{author}{{Micela}, G.},
  \bibinfo{author}{{Sciortino}, S.}, \bibinfo{author}{{Cody}, A.M.},
  \bibinfo{author}{{Guarcello}, M.G.}, \bibinfo{author}{{Morales-Calderòn},
  M.}, \bibinfo{author}{{Rebull}, L.}, \bibinfo{author}{{Stauffer}, J.R.},
  \bibinfo{year}{2018}.
\newblock \bibinfo{title}{{A multi-wavelength view of magnetic flaring from PMS
  stars}}.
\newblock \bibinfo{journal}{A\&A} \bibinfo{volume}{620}, \bibinfo{pages}{55}.
%Type = Article
\bibitem[{{Fuhrmeister} et~al.(2011){Fuhrmeister}, {Lalitha}, {Poppenhaeger},
  {Rudolf}, {Liefke}, {Reiners}, {Schmitt} and {Ness}}]{Fuhrmeister:2011}
\bibinfo{author}{{Fuhrmeister}, B.}, \bibinfo{author}{{Lalitha}, S.},
  \bibinfo{author}{{Poppenhaeger}, K.}, \bibinfo{author}{{Rudolf}, N.},
  \bibinfo{author}{{Liefke}, C.}, \bibinfo{author}{{Reiners}, A.},
  \bibinfo{author}{{Schmitt}, J.H.M.M.}, \bibinfo{author}{{Ness}, J.U.},
  \bibinfo{year}{2011}.
\newblock \bibinfo{title}{{Multi-wavelength observations of Proxima Centauri}}.
\newblock \bibinfo{journal}{A\&A} \bibinfo{volume}{534}, \bibinfo{pages}{133}.
%Type = Article
\bibitem[{{Gaia Collaboration} et~al.(2018){Gaia Collaboration}, {Brown},
  {Vallenari}, {Prusti}, {de Bruijne}, {Babusiaux}, {Bailer-Jones}, {Biermann},
  {Evans}, {Eyer}, {Jansen} et~al.}]{Gaia:2018}
\bibinfo{author}{{Gaia Collaboration}}, \bibinfo{author}{{Brown}, A.G.A.},
  \bibinfo{author}{{Vallenari}, A.}, \bibinfo{author}{{Prusti}, T.},
  \bibinfo{author}{{de Bruijne}, J.H.J.}, \bibinfo{author}{{Babusiaux}, C.},
  \bibinfo{author}{{Bailer-Jones}, C.A.L.}, \bibinfo{author}{{Biermann}, M.},
  \bibinfo{author}{{Evans}, D.W.}, \bibinfo{author}{{Eyer}, L.},
  \bibinfo{author}{{Jansen}, F.}, et~al., \bibinfo{year}{2018}.
\newblock \bibinfo{title}{{Gaia Data Release 2. Summary of the contents and
  survey properties}}.
\newblock \bibinfo{journal}{A\&A} \bibinfo{volume}{616}, \bibinfo{pages}{A1}.
%Type = Article
\bibitem[{{Garcia} et~al.(1980){Garcia}, {Baliunas}, {Conroy}, {Johnston},
  {Ralph}, {Roberts}, {Schwartz} and {Tonry}}]{Garcia:1980}
\bibinfo{author}{{Garcia}, M.}, \bibinfo{author}{{Baliunas}, S.L.},
  \bibinfo{author}{{Conroy}, M.}, \bibinfo{author}{{Johnston}, M.D.},
  \bibinfo{author}{{Ralph}, E.}, \bibinfo{author}{{Roberts}, W.},
  \bibinfo{author}{{Schwartz}, D.A.}, \bibinfo{author}{{Tonry}, J.},
  \bibinfo{year}{1980}.
\newblock \bibinfo{title}{{Optical identification of H 0123+07.5 and 4U 1137-65
  : hard X-ray emission from RS CVn systems.}}
\newblock \bibinfo{journal}{ApJ} \bibinfo{volume}{Vol. 240},
  \bibinfo{pages}{107}.
%Type = Article
\bibitem[{{Guarcello} et~al.(2019){Guarcello}, {Micela}, {Sciortino},
  {López-Santiago}, {Argiroffi}, {Reale}, {Flaccomio}, {Alvarado-Gómez},
  {Antoniou} et~al.}]{Guarcello:2019}
\bibinfo{author}{{Guarcello}, M.G.}, \bibinfo{author}{{Micela}, G.},
  \bibinfo{author}{{Sciortino}, S.}, \bibinfo{author}{{López-Santiago}, J.},
  \bibinfo{author}{{Argiroffi}, C.}, \bibinfo{author}{{Reale}, F.},
  \bibinfo{author}{{Flaccomio}, E.}, \bibinfo{author}{{Alvarado-Gómez}, J.D.},
  \bibinfo{author}{{Antoniou}, V.}, et~al., \bibinfo{year}{2019}.
\newblock \bibinfo{title}{{Simultaneous Kepler/K2 and XMM-Newton observations
  of superflares in the Pleiades}}.
\newblock \bibinfo{journal}{A\&A} \bibinfo{volume}{622}, \bibinfo{pages}{210}.
%Type = Article
\bibitem[{{Haisch}(1983)}]{Haisch:1983}
\bibinfo{author}{{Haisch}, B.M.}, \bibinfo{year}{1983}.
\newblock \bibinfo{title}{{X-ray observations of stellar flares}}.
\newblock \bibinfo{journal}{in IAU Colloq. 71, Activity in Red-Dwarf Stars, ed.
  P. B. Byrne \& M. Rodono (Dordrecht: Reidel)} , \bibinfo{pages}{255}.
%Type = Article
\bibitem[{{Hummel} et~al.(2017){Hummel}, {Monnier}, {Roettenbacher}, {Torres},
  {Henry}, {Korhonen}, {Beasley}, {Schaefer}, {Turner} and {Ten
  Brummelaar}}]{Hummel:2017}
\bibinfo{author}{{Hummel}, C.A.}, \bibinfo{author}{{Monnier}, J.D.},
  \bibinfo{author}{{Roettenbacher}, R.M.}, \bibinfo{author}{{Torres}, G.},
  \bibinfo{author}{{Henry}, G.W.}, \bibinfo{author}{{Korhonen}, H.},
  \bibinfo{author}{{Beasley}, A.}, \bibinfo{author}{{Schaefer}, G.H.},
  \bibinfo{author}{{Turner}, N.H.}, \bibinfo{author}{{Ten Brummelaar}, T.},
  \bibinfo{year}{2017}.
\newblock \bibinfo{title}{{Orbital Elements and Stellar Parameters of the
  Active Binary UX Arietis}}.
\newblock \bibinfo{journal}{ApJ} \bibinfo{volume}{844}, \bibinfo{pages}{115}.
%Type = Article
\bibitem[{{Jing} et~al.(2016){Jing}, {Xu}, {Cao}, {Liu}, {Gary} and
  {Wang}}]{Jing:2016}
\bibinfo{author}{{Jing}, J.}, \bibinfo{author}{{Xu}, Y.},
  \bibinfo{author}{{Cao}, W.}, \bibinfo{author}{{Liu}, C.},
  \bibinfo{author}{{Gary}, D.}, \bibinfo{author}{{Wang}, H.},
  \bibinfo{year}{2016}.
\newblock \bibinfo{title}{{Unprecedented Fine Structure of a Solar Flare
  Revealed by the 1.6 m New Solar Telescope}}.
\newblock \bibinfo{journal}{NatSR} \bibinfo{volume}{6}, \bibinfo{pages}{24319}.
%Type = Article
\bibitem[{{Johns-Krull} et~al.(1997){Johns-Krull}, {Hawley}, {Basri} and
  {Valenti}}]{Johns-Krull:1997}
\bibinfo{author}{{Johns-Krull}, C.M.}, \bibinfo{author}{{Hawley}, S.L.},
  \bibinfo{author}{{Basri}, G.}, \bibinfo{author}{{Valenti}, J.A.},
  \bibinfo{year}{1997}.
\newblock \bibinfo{title}{{Hamilton Echelle Spectroscopy of the 1993 March 6
  Solar Flare}}.
\newblock \bibinfo{journal}{ApJS} \bibinfo{volume}{112}, \bibinfo{pages}{221}.
%Type = Article
\bibitem[{{Kahler} et~al.(1982){Kahler}, {Golub}, {Harnden}, {Liller},
  {Seward}, {Vaiana}, {Lovell}, {Davis}, {Spencer}, {Whitehouse}, {Feldman},
  {Viner}, {Leslie}, {Kahn}, {Mason}, {Davis}, {Crannell}, {Hobbs},
  {Schneeberger}, {Worden}, {Schommer}, {Vogt}, {Pettersen}, {Coleman},
  {Karpen}, {Giampapa}, {Hege}, {Pazzani}, {Rodono}, {Romeo} and
  {Chugainov}}]{Kahler:1982}
\bibinfo{author}{{Kahler}, S.}, \bibinfo{author}{{Golub}, L.},
  \bibinfo{author}{{Harnden}, F.R.}, \bibinfo{author}{{Liller}, W.},
  \bibinfo{author}{{Seward}, F.}, \bibinfo{author}{{Vaiana}, G.},
  \bibinfo{author}{{Lovell}, B.}, \bibinfo{author}{{Davis}, R.J.},
  \bibinfo{author}{{Spencer}, R.E.}, \bibinfo{author}{{Whitehouse}, D.R.},
  \bibinfo{author}{{Feldman}, P.A.}, \bibinfo{author}{{Viner}, M.R.},
  \bibinfo{author}{{Leslie}, B.}, \bibinfo{author}{{Kahn}, S.M.},
  \bibinfo{author}{{Mason}, K.O.}, \bibinfo{author}{{Davis}, M.M.},
  \bibinfo{author}{{Crannell}, C.J.}, \bibinfo{author}{{Hobbs}, R.W.},
  \bibinfo{author}{{Schneeberger}, T.J.}, \bibinfo{author}{{Worden}, S.P.},
  \bibinfo{author}{{Schommer}, R.A.}, \bibinfo{author}{{Vogt}, S.S.},
  \bibinfo{author}{{Pettersen}, B.R.}, \bibinfo{author}{{Coleman}, G.D.},
  \bibinfo{author}{{Karpen}, J.T.}, \bibinfo{author}{{Giampapa}, M.S.},
  \bibinfo{author}{{Hege}, E.K.}, \bibinfo{author}{{Pazzani}, V.},
  \bibinfo{author}{{Rodono}, M.}, \bibinfo{author}{{Romeo}, G.},
  \bibinfo{author}{{Chugainov}, P.F.}, \bibinfo{year}{1982}.
\newblock \bibinfo{title}{{Coordinated X-ray, optical and radio observations of
  flaring activityon YZ Canis Minoris.}}
\newblock \bibinfo{journal}{ApJ} \bibinfo{volume}{252}, \bibinfo{pages}{239}.
%Type = Article
\bibitem[{{Kashapova} et~al.(2021){Kashapova}, {Broomhall}, {Larionova},
  {Kupriyanova} and {Motyk}}]{Kashapova:2021}
\bibinfo{author}{{Kashapova}, L.K.}, \bibinfo{author}{{Broomhall}, A.},
  \bibinfo{author}{{Larionova}, A.I.}, \bibinfo{author}{{Kupriyanova}, E.G.},
  \bibinfo{author}{{Motyk}, I.D.}, \bibinfo{year}{2021}.
\newblock \bibinfo{title}{{The morphology of average solar flare time profiles
  from observations of the Sun's lower atmosphere}}.
\newblock \bibinfo{journal}{MNRAS} \bibinfo{volume}{502},
  \bibinfo{pages}{3922}.
%Type = Article
\bibitem[{{Kashyap} and {Drake}(1999)}]{Kashyap:1999}
\bibinfo{author}{{Kashyap}, V.}, \bibinfo{author}{{Drake}, J.J.},
  \bibinfo{year}{1999}.
\newblock \bibinfo{title}{{On X-Ray Variability in Active Binary Stars}}.
\newblock \bibinfo{journal}{ApJ} \bibinfo{volume}{524},
  \bibinfo{pages}{988--999}.
%Type = Article
\bibitem[{{Kawagoe} et~al.(2014){Kawagoe}, {Tsuboi}, {Negoro}, {Kawai},
  {Sakamoto}, {Nakahira}, {Serino}, {Ueno}, {Tomida}, {Kimura}, {Nakagawa},
  {Mihara}, {Sugizaki}, {Morii}, {Sugimoto}, {Takagi}, {Yoshikawa}, {Matsuoka},
  {Usui}, {Yoshii} et~al.}]{Kawagoe:2014}
\bibinfo{author}{{Kawagoe}, A.}, \bibinfo{author}{{Tsuboi}, Y.},
  \bibinfo{author}{{Negoro}, H.}, \bibinfo{author}{{Kawai}, N.},
  \bibinfo{author}{{Sakamoto}, T.}, \bibinfo{author}{{Nakahira}, S.},
  \bibinfo{author}{{Serino}, M.}, \bibinfo{author}{{Ueno}, S.},
  \bibinfo{author}{{Tomida}, H.}, \bibinfo{author}{{Kimura}, M.},
  \bibinfo{author}{{Nakagawa}, Y.E.}, \bibinfo{author}{{Mihara}, T.},
  \bibinfo{author}{{Sugizaki}, M.}, \bibinfo{author}{{Morii}, M.},
  \bibinfo{author}{{Sugimoto}, J.}, \bibinfo{author}{{Takagi}, T.},
  \bibinfo{author}{{Yoshikawa}, A.}, \bibinfo{author}{{Matsuoka}, M.},
  \bibinfo{author}{{Usui}, R.}, \bibinfo{author}{{Yoshii}, T.}, et~al.,
  \bibinfo{year}{2014}.
\newblock \bibinfo{title}{{MAXI/GSC detection of a big flare from UX Ari}}.
\newblock \bibinfo{journal}{ATel} \bibinfo{volume}{6315}.
%Type = Article
\bibitem[{{Kontar} et~al.(2011){Kontar}, {Hannah} and {Bian}}]{Kontar:2011}
\bibinfo{author}{{Kontar}, E.P.}, \bibinfo{author}{{Hannah}, I.G.},
  \bibinfo{author}{{Bian}, N.H.}, \bibinfo{year}{2011}.
\newblock \bibinfo{title}{{Acceleration, Magnetic Fluctuations, and Cross-field
  Transport of Energetic Electrons in a Solar Flare Loop}}.
\newblock \bibinfo{journal}{ApJL} \bibinfo{volume}{730}, \bibinfo{pages}{22}.
%Type = Article
\bibitem[{{Kuznetsov} and {Kolotkov}(2021)}]{Kuznetsov:2021}
\bibinfo{author}{{Kuznetsov}, A.A.}, \bibinfo{author}{{Kolotkov}, D.Y.},
  \bibinfo{year}{2021}.
\newblock \bibinfo{title}{{Stellar Superflares Observed Simultaneously with
  Kepler and XMM-Newton}}.
\newblock \bibinfo{journal}{ApJ} \bibinfo{volume}{912}, \bibinfo{pages}{81}.
%Type = Article
\bibitem[{{Massi} et~al.(1998){Massi}, {Neidhofer}, {Torricelli-Ciamponi} and
  {Chiuderi-Drago}}]{Massi:1998}
\bibinfo{author}{{Massi}, M.}, \bibinfo{author}{{Neidhofer}, J.},
  \bibinfo{author}{{Torricelli-Ciamponi}, G.},
  \bibinfo{author}{{Chiuderi-Drago}, F.}, \bibinfo{year}{1998}.
\newblock \bibinfo{title}{{Activity cycles in UX ARIETIS}}.
\newblock \bibinfo{journal}{A\&A} \bibinfo{volume}{332}, \bibinfo{pages}{149}.
%Type = Article
\bibitem[{{Massi} et~al.(2005){Massi}, {Neidhöfer}, {Carpentier} and
  {Ros}}]{Massi:2005}
\bibinfo{author}{{Massi}, M.}, \bibinfo{author}{{Neidhöfer}, J.},
  \bibinfo{author}{{Carpentier}, Y.}, \bibinfo{author}{{Ros}, E.},
  \bibinfo{year}{2005}.
\newblock \bibinfo{title}{{Discovery of Solar Rieger periodicities in another
  star}}.
\newblock \bibinfo{journal}{A\&A} \bibinfo{volume}{435}, \bibinfo{pages}{1}.
%Type = Article
\bibitem[{{Massi} and {Ros}(2002)}]{Massi:2002}
\bibinfo{author}{{Massi}, M.}, \bibinfo{author}{{Ros}, E.},
  \bibinfo{year}{2002}.
\newblock \bibinfo{title}{{Investigation of magnetic loop structures in the
  corona of UX Arietis}}.
\newblock \bibinfo{journal}{6th European VLBI Network Symposium on New
  Developments in VLBI Science and Technology} , \bibinfo{pages}{275}.
%Type = Article
\bibitem[{{Matsumura} et~al.(2011){Matsumura}, {Tsuboi}, {Yamazaki}, {Morii},
  {Ueno}, {Tomida}, {Kohama}, {Ishikawa}, {Mihara}, {Sugizaki}, {Serino},
  {Nakahira}, {Yamamoto}, {Sootome}, {Matsuoka}, {Kawai}, {Sugimori}, {Usui},
  {Toizumi}, {Yoshida} et~al.}]{Matsumura:2011}
\bibinfo{author}{{Matsumura}, T.}, \bibinfo{author}{{Tsuboi}, Y.},
  \bibinfo{author}{{Yamazaki}, K.}, \bibinfo{author}{{Morii}, M.},
  \bibinfo{author}{{Ueno}, S.}, \bibinfo{author}{{Tomida}, H.},
  \bibinfo{author}{{Kohama}, M.}, \bibinfo{author}{{Ishikawa}, M.},
  \bibinfo{author}{{Mihara}, T.}, \bibinfo{author}{{Sugizaki}, M.},
  \bibinfo{author}{{Serino}, M.}, \bibinfo{author}{{Nakahira}, S.},
  \bibinfo{author}{{Yamamoto}, T.}, \bibinfo{author}{{Sootome}, T.},
  \bibinfo{author}{{Matsuoka}, M.}, \bibinfo{author}{{Kawai}, N.},
  \bibinfo{author}{{Sugimori}, K.}, \bibinfo{author}{{Usui}, R.},
  \bibinfo{author}{{Toizumi}, T.}, \bibinfo{author}{{Yoshida}, A.}, et~al.,
  \bibinfo{year}{2011}.
\newblock \bibinfo{title}{{MAXI/GSC detection of an active state of the RS CVn
  type star UX Ari}}.
\newblock \bibinfo{journal}{ATel} \bibinfo{volume}{3308}.
%Type = Article
\bibitem[{{Matsuoka} et~al.(2009){Matsuoka}, {Kawasaki}, {Ueno}, {Tomida},
  {Kohama}, {Suzuki}, {Adachi}, {Ishikawa}, {Mihara}, {Sugizaki}, {Isobe},
  {Nakagawa}, {Tsunemi}, {Miyata}, {Kawai}, {Kataoka}, {Morii}, {Yoshida},
  {Negoro}, {Nakajima}, {Ueda}, {Chujo}, {Yamaoka}, {Yamazaki}, {Nakahira},
  {You}, {Ishiwata}, {Miyoshi}, {Eguchi}, {Hiroi}, {Katayama} and
  {Ebisawa}}]{Matsuoka:2009}
\bibinfo{author}{{Matsuoka}, M.}, \bibinfo{author}{{Kawasaki}, K.},
  \bibinfo{author}{{Ueno}, S.}, \bibinfo{author}{{Tomida}, H.},
  \bibinfo{author}{{Kohama}, M.}, \bibinfo{author}{{Suzuki}, M.},
  \bibinfo{author}{{Adachi}, Y.}, \bibinfo{author}{{Ishikawa}, M.},
  \bibinfo{author}{{Mihara}, T.}, \bibinfo{author}{{Sugizaki}, M.},
  \bibinfo{author}{{Isobe}, N.}, \bibinfo{author}{{Nakagawa}, Y.},
  \bibinfo{author}{{Tsunemi}, H.}, \bibinfo{author}{{Miyata}, E.},
  \bibinfo{author}{{Kawai}, N.}, \bibinfo{author}{{Kataoka}, J.},
  \bibinfo{author}{{Morii}, M.}, \bibinfo{author}{{Yoshida}, A.},
  \bibinfo{author}{{Negoro}, H.}, \bibinfo{author}{{Nakajima}, M.},
  \bibinfo{author}{{Ueda}, Y.}, \bibinfo{author}{{Chujo}, H.},
  \bibinfo{author}{{Yamaoka}, K.}, \bibinfo{author}{{Yamazaki}, O.},
  \bibinfo{author}{{Nakahira}, S.}, \bibinfo{author}{{You}, T.},
  \bibinfo{author}{{Ishiwata}, R.}, \bibinfo{author}{{Miyoshi}, S.},
  \bibinfo{author}{{Eguchi}, S.}, \bibinfo{author}{{Hiroi}, K.},
  \bibinfo{author}{{Katayama}, H.}, \bibinfo{author}{{Ebisawa}, K.},
  \bibinfo{year}{2009}.
\newblock \bibinfo{title}{{The MAXI Mission on the ISS: Science and Instruments
  for Monitoring All-Sky X-Ray Images}}.
\newblock \bibinfo{journal}{PASJ} \bibinfo{volume}{61}, \bibinfo{pages}{999}.
\newblock \href{http://arxiv.org/abs/0906.0631}{{\tt arXiv:0906.0631}}.
%Type = Article
\bibitem[{{Mihara} et~al.(2011){Mihara}, {Nakajima}, {Sugizaki}, {Serino},
  {Matsuoka}, {Kohama}, {Kawasaki}, {Tomida}, {Ueno}, {Kawai}, {Kataoka},
  {Morii}, {Yoshida}, {Yamaoka}, {Nakahira}, {Negoro}, {Isobe}, {Yamauchi} and
  {Sakurai}}]{Mihara:2011}
\bibinfo{author}{{Mihara}, T.}, \bibinfo{author}{{Nakajima}, M.},
  \bibinfo{author}{{Sugizaki}, M.}, \bibinfo{author}{{Serino}, M.},
  \bibinfo{author}{{Matsuoka}, M.}, \bibinfo{author}{{Kohama}, M.},
  \bibinfo{author}{{Kawasaki}, K.}, \bibinfo{author}{{Tomida}, H.},
  \bibinfo{author}{{Ueno}, S.}, \bibinfo{author}{{Kawai}, N.},
  \bibinfo{author}{{Kataoka}, J.}, \bibinfo{author}{{Morii}, M.},
  \bibinfo{author}{{Yoshida}, A.}, \bibinfo{author}{{Yamaoka}, K.},
  \bibinfo{author}{{Nakahira}, S.}, \bibinfo{author}{{Negoro}, H.},
  \bibinfo{author}{{Isobe}, N.}, \bibinfo{author}{{Yamauchi}, M.},
  \bibinfo{author}{{Sakurai}, I.}, \bibinfo{year}{2011}.
\newblock \bibinfo{title}{{Gas Slit Camera (GSC) onboard MAXI on ISS}}.
\newblock \bibinfo{journal}{PASJ} \bibinfo{volume}{63},
  \bibinfo{pages}{623--634}.
\newblock \href{http://arxiv.org/abs/1103.4224}{{\tt arXiv:1103.4224}}.
%Type = Article
\bibitem[{{Nakahira} et~al.(2013){Nakahira}, {Ebisawa} and
  {Negoro}}]{Nakahira:2013}
\bibinfo{author}{{Nakahira}, S.}, \bibinfo{author}{{Ebisawa}, K.},
  \bibinfo{author}{{Negoro}, H.}, \bibinfo{year}{2013}.
\newblock \bibinfo{title}{{Development of the MAXI/GSC all-sky data archive
  system}}.
\newblock \bibinfo{journal}{JAXA Research and development report: Journal of
  Space Science Informatics Japan} \bibinfo{volume}{2}, \bibinfo{pages}{29}.
%Type = Article
\bibitem[{{Nakamura} et~al.(2016){Nakamura}, {Tsuboi}, {Sasaki}, {Serino},
  {Nakahira}, {Ueno}, {Tomida}, {Ishikawa}, {Nakagawa}, {Sugawara}, {Mihara},
  {Sugizaki}, {Iwakiri}, {Shidatsu}, {Sugimoto}, {Takagi}, {Matsuoka}, {Kawai},
  {Isobe}, {Sugita} et~al.}]{Nakamura:2016}
\bibinfo{author}{{Nakamura}, Y.}, \bibinfo{author}{{Tsuboi}, Y.},
  \bibinfo{author}{{Sasaki}, R.}, \bibinfo{author}{{Serino}, M.},
  \bibinfo{author}{{Nakahira}, S.}, \bibinfo{author}{{Ueno}, S.},
  \bibinfo{author}{{Tomida}, H.}, \bibinfo{author}{{Ishikawa}, M.},
  \bibinfo{author}{{Nakagawa}, Y.E.}, \bibinfo{author}{{Sugawara}, Y.},
  \bibinfo{author}{{Mihara}, T.}, \bibinfo{author}{{Sugizaki}, M.},
  \bibinfo{author}{{Iwakiri}, W.}, \bibinfo{author}{{Shidatsu}, M.},
  \bibinfo{author}{{Sugimoto}, J.}, \bibinfo{author}{{Takagi}, T.},
  \bibinfo{author}{{Matsuoka}, M.}, \bibinfo{author}{{Kawai}, N.},
  \bibinfo{author}{{Isobe}, N.}, \bibinfo{author}{{Sugita}, S.}, et~al.,
  \bibinfo{year}{2016}.
\newblock \bibinfo{title}{{MAXI/GSC detection of a possible X-ray flare from an
  RS CVn star AR Psc}}.
\newblock \bibinfo{journal}{ATel} \bibinfo{volume}{9288}.
%Type = Article
\bibitem[{{Namekata} et~al.(2020){Namekata}, {Maehara}, {Sasaki}, {Kawai},
  {Notsu}, {Kowalski}, {Allred}, {Iwakiri} et~al.}]{Namekata:2020}
\bibinfo{author}{{Namekata}, K.}, \bibinfo{author}{{Maehara}, H.},
  \bibinfo{author}{{Sasaki}, R.}, \bibinfo{author}{{Kawai}, H.},
  \bibinfo{author}{{Notsu}, Y.}, \bibinfo{author}{{Kowalski}, A.F.},
  \bibinfo{author}{{Allred}, J.C.}, \bibinfo{author}{{Iwakiri}, W.}, et~al.,
  \bibinfo{year}{2020}.
\newblock \bibinfo{title}{{Optical and X-ray observations of stellar flares on
  an active M dwarf AD Leonis with the Seimei Telescope, SCAT, NICER, and
  OISTER}}.
\newblock \bibinfo{journal}{PASJ} \bibinfo{volume}{72}, \bibinfo{pages}{68}.
%Type = Article
\bibitem[{{Negoro} et~al.(2016){Negoro}, {Kohama}, {Serino}, {Saito},
  {Takahashi}, {Miyoshi}, {Ozawa}, {Suwa}, {Asada}, {Fukushima}, {Eguchi},
  {Kawai}, {Kennea}, {Mihara}, {Morii}, {Nakahira}, {Ogawa}, {Sugawara},
  {Tomida}, {Ueno}, {Ishikawa}, {Isobe}, {Kawamuro}, {Kimura}, {Masumitsu},
  {Nakagawa}, {Nakajima}, {Sakamoto}, {Shidatsu}, {Sugizaki}, {Sugimoto},
  {Suzuki}, {Takagi}, {Tanaka}, {Tsuboi}, {Tsunemi}, {Ueda}, {Yamaoka},
  {Yamauchi}, {Yoshida} and {Matsuoka}}]{Negoro:2016}
\bibinfo{author}{{Negoro}, H.}, \bibinfo{author}{{Kohama}, M.},
  \bibinfo{author}{{Serino}, M.}, \bibinfo{author}{{Saito}, H.},
  \bibinfo{author}{{Takahashi}, T.}, \bibinfo{author}{{Miyoshi}, S.},
  \bibinfo{author}{{Ozawa}, H.}, \bibinfo{author}{{Suwa}, F.},
  \bibinfo{author}{{Asada}, M.}, \bibinfo{author}{{Fukushima}, K.},
  \bibinfo{author}{{Eguchi}, S.}, \bibinfo{author}{{Kawai}, N.},
  \bibinfo{author}{{Kennea}, J.}, \bibinfo{author}{{Mihara}, T.},
  \bibinfo{author}{{Morii}, M.}, \bibinfo{author}{{Nakahira}, S.},
  \bibinfo{author}{{Ogawa}, Y.}, \bibinfo{author}{{Sugawara}, A.},
  \bibinfo{author}{{Tomida}, H.}, \bibinfo{author}{{Ueno}, S.},
  \bibinfo{author}{{Ishikawa}, M.}, \bibinfo{author}{{Isobe}, N.},
  \bibinfo{author}{{Kawamuro}, T.}, \bibinfo{author}{{Kimura}, M.},
  \bibinfo{author}{{Masumitsu}, T.}, \bibinfo{author}{{Nakagawa}, Y.E.},
  \bibinfo{author}{{Nakajima}, M.}, \bibinfo{author}{{Sakamoto}, T.},
  \bibinfo{author}{{Shidatsu}, M.}, \bibinfo{author}{{Sugizaki}, M.},
  \bibinfo{author}{{Sugimoto}, J.}, \bibinfo{author}{{Suzuki}, K.},
  \bibinfo{author}{{Takagi}, T.}, \bibinfo{author}{{Tanaka}, K.},
  \bibinfo{author}{{Tsuboi}, Y.}, \bibinfo{author}{{Tsunemi}, H.},
  \bibinfo{author}{{Ueda}, Y.}, \bibinfo{author}{{Yamaoka}, K.},
  \bibinfo{author}{{Yamauchi}, M.}, \bibinfo{author}{{Yoshida}, A.},
  \bibinfo{author}{{Matsuoka}, M.}, \bibinfo{year}{2016}.
\newblock \bibinfo{title}{{The MAXI/GSC Nova-Alert System and results of its
  first 68 months}}.
\newblock \bibinfo{journal}{PASJ} \bibinfo{volume}{68}, \bibinfo{pages}{1}.
%Type = Article
\bibitem[{{Peterson} et~al.(2011){Peterson}, {Mutel}, {Lestrade}, {Güdel} and
  {Goss}}]{Peterson:2011}
\bibinfo{author}{{Peterson}, W.M.}, \bibinfo{author}{{Mutel}, R.L.},
  \bibinfo{author}{{Lestrade}, J.F.}, \bibinfo{author}{{Güdel}, M.},
  \bibinfo{author}{{Goss}, W.M.}, \bibinfo{year}{2011}.
\newblock \bibinfo{title}{{Radio Astrometry of the Triple Systems Algol and UX
  Arietis}}.
\newblock \bibinfo{journal}{ApJ} \bibinfo{volume}{737}, \bibinfo{pages}{104}.
%Type = Article
\bibitem[{{Sasaki} et~al.(2021){Sasaki}, {Tsuboi}, {Iwakiri}, {Nakahira},
  {Maeda}, {Gendreau}, {Corcoran}, {Hamaguchi}, {Arzoumanian}, {Markwardt},
  {Enoto}, {Sato}, {Kawai}, {Mihara}, {Shidatsu}, {Negoro} and
  {Serino}}]{Sasaki:2021}
\bibinfo{author}{{Sasaki}, R.}, \bibinfo{author}{{Tsuboi}, Y.},
  \bibinfo{author}{{Iwakiri}, W.}, \bibinfo{author}{{Nakahira}, S.},
  \bibinfo{author}{{Maeda}, Y.}, \bibinfo{author}{{Gendreau}, K.},
  \bibinfo{author}{{Corcoran}, M.F.}, \bibinfo{author}{{Hamaguchi}, K.},
  \bibinfo{author}{{Arzoumanian}, Z.}, \bibinfo{author}{{Markwardt}, C.B.},
  \bibinfo{author}{{Enoto}, T.}, \bibinfo{author}{{Sato}, T.},
  \bibinfo{author}{{Kawai}, H.}, \bibinfo{author}{{Mihara}, T.},
  \bibinfo{author}{{Shidatsu}, M.}, \bibinfo{author}{{Negoro}, H.},
  \bibinfo{author}{{Serino}, M.}, \bibinfo{year}{2021}.
\newblock \bibinfo{title}{{The RS CVn-type Star GT Mus Shows Most Energetic
  X-Ray Flares Throughout the 2010s}}.
\newblock \bibinfo{journal}{ApJ} \bibinfo{volume}{910}, \bibinfo{pages}{25}.
\newblock \href{http://arxiv.org/abs/2103.16822}{{\tt arXiv:2103.16822}}.
%Type = Article
\bibitem[{{Sasaki} et~al.(2017){Sasaki}, {Tsuboi}, {Katsuda}, {Yabuki},
  {Nakamura}, {Sugawara}, {Matsuoka} and Team}]{Sasaki:2017}
\bibinfo{author}{{Sasaki}, R.}, \bibinfo{author}{{Tsuboi}, Y.},
  \bibinfo{author}{{Katsuda}, S.}, \bibinfo{author}{{Yabuki}, K.},
  \bibinfo{author}{{Nakamura}, Y.}, \bibinfo{author}{{Sugawara}, Y.},
  \bibinfo{author}{{Matsuoka}, M.}, \bibinfo{author}{Team, M.},
  \bibinfo{year}{2017}.
\newblock \bibinfo{title}{{A Correlation on Stellar Flares detected with MAXI
  Quiescent Luminosity vs. Flare Energy}}.
\newblock \bibinfo{journal}{The X-ray Universe 2017, Proceedings of the
  conference held 6-9 June, 2017 in Rome, Italy. Edited by J.-U. Ness and S.
  Migliari.} \bibinfo{volume}{201}.
%Type = Article
\bibitem[{{Shan} et~al.(2006){Shan}, {Liu} and {Gu}}]{Shan:2006}
\bibinfo{author}{{Shan}, H.}, \bibinfo{author}{{Liu}, X.},
  \bibinfo{author}{{Gu}, S.}, \bibinfo{year}{2006}.
\newblock \bibinfo{title}{{The atmospheric parameters, abundances and magnetic
  field of the AR piscium primary}}.
\newblock \bibinfo{journal}{New Astron.} \bibinfo{volume}{11},
  \bibinfo{pages}{287--292}.
%Type = Article
\bibitem[{{Shibata} and {Magara}(2011)}]{Shibata:2011}
\bibinfo{author}{{Shibata}, K.}, \bibinfo{author}{{Magara}, T.},
  \bibinfo{year}{2011}.
\newblock \bibinfo{title}{{Solar Flares: Magnetohydrodynamic Processes}}.
\newblock \bibinfo{journal}{Living Rev. Sol. Phys.} \bibinfo{volume}{8},
  \bibinfo{pages}{6}.
%Type = Article
\bibitem[{{Shibata} and {Yokoyama}(1999)}]{Shibata:1999}
\bibinfo{author}{{Shibata}, K.}, \bibinfo{author}{{Yokoyama}, T.},
  \bibinfo{year}{1999}.
\newblock \bibinfo{title}{{Origin of the Universal Correlation between the
  Flare Temperature and the Emission Measure for Solar and Stellar Flares}}.
\newblock \bibinfo{journal}{ApJ} \bibinfo{volume}{526}, \bibinfo{pages}{L49}.
%Type = Article
\bibitem[{{Smith} et~al.(2001){Smith}, {Brickhouse}, {Liedahl} and
  {Raymond}}]{Smith:2001}
\bibinfo{author}{{Smith}, R.K.}, \bibinfo{author}{{Brickhouse}, N.S.},
  \bibinfo{author}{{Liedahl}, D.A.}, \bibinfo{author}{{Raymond}, J.C.},
  \bibinfo{year}{2001}.
\newblock \bibinfo{title}{{Collisional Plasma Models with APEC/APED:
  Emission-Line Diagnostics of Hydrogen-like and Helium-like Ions}}.
\newblock \bibinfo{journal}{ApJ} \bibinfo{volume}{556}, \bibinfo{pages}{91}.
\newblock \href{http://arxiv.org/abs/astro-ph/0106478}{{\tt
  arXiv:astro-ph/0106478}}.
%Type = Article
\bibitem[{{Thomas} and {Teske}(1971)}]{Thomas:1971}
\bibinfo{author}{{Thomas}, R.J.}, \bibinfo{author}{{Teske}, R.G.},
  \bibinfo{year}{1971}.
\newblock \bibinfo{title}{{Solar Soft X-Rays and Solar Activity. II: Soft X-Ray
  Emission during Solar Flares}}.
\newblock \bibinfo{journal}{Sol. Phys.} \bibinfo{volume}{16},
  \bibinfo{pages}{431}.
%Type = Article
\bibitem[{{Tsuboi} et~al.(1998){Tsuboi}, {Koyama}, {Murakami}, {Hayashi},
  {Skinner} and {Ueno}}]{Tsuboi:1998}
\bibinfo{author}{{Tsuboi}, Y.}, \bibinfo{author}{{Koyama}, K.},
  \bibinfo{author}{{Murakami}, H.}, \bibinfo{author}{{Hayashi}, M.},
  \bibinfo{author}{{Skinner}, S.}, \bibinfo{author}{{Ueno}, S.},
  \bibinfo{year}{1998}.
\newblock \bibinfo{title}{{ASCA Detection of a Superhot 100 Million K X-Ray
  Flare on the Weak-Lined T Tauri Star V773 Tauri}}.
\newblock \bibinfo{journal}{ApJ} \bibinfo{volume}{503}, \bibinfo{pages}{894}.
%Type = Article
\bibitem[{{Tsuboi} et~al.(2016){Tsuboi}, {Yamazaki}, {Sugawara}, {Kawagoe},
  {Kaneto}, {Iizuka}, {Matsumura}, {Nakahira} et~al.}]{Tsuboi:2016}
\bibinfo{author}{{Tsuboi}, Y.}, \bibinfo{author}{{Yamazaki}, K.},
  \bibinfo{author}{{Sugawara}, Y.}, \bibinfo{author}{{Kawagoe}, A.},
  \bibinfo{author}{{Kaneto}, S.}, \bibinfo{author}{{Iizuka}, R.},
  \bibinfo{author}{{Matsumura}, T.}, \bibinfo{author}{{Nakahira}, S.}, et~al.,
  \bibinfo{year}{2016}.
\newblock \bibinfo{title}{{Large X-ray flares on stars detected with MAXI/GSC:
  A universal correlation between the duration of a flare and its X-ray
  luminosity}}.
\newblock \bibinfo{journal}{PASJ} \bibinfo{volume}{68}, \bibinfo{pages}{90}.
\newblock \href{http://arxiv.org/abs/1609.01925}{{\tt arXiv:1609.01925}}.
%Type = Article
\bibitem[{{Tsuru} et~al.(1989){Tsuru}, {Makishima}, {Ohashi}, {Inoue},
  {Koyama}, {Turner}, {Barstow}, {McHardy}, {Pye}, {Tsunemi}, {Kitamoto},
  {Taylor} and {Nelson}}]{Tsuru:1989}
\bibinfo{author}{{Tsuru}, T.}, \bibinfo{author}{{Makishima}, K.},
  \bibinfo{author}{{Ohashi}, T.}, \bibinfo{author}{{Inoue}, H.},
  \bibinfo{author}{{Koyama}, K.}, \bibinfo{author}{{Turner}, M.J.L.},
  \bibinfo{author}{{Barstow}, M.A.}, \bibinfo{author}{{McHardy}, I.M.},
  \bibinfo{author}{{Pye}, J.P.}, \bibinfo{author}{{Tsunemi}, H.},
  \bibinfo{author}{{Kitamoto}, S.}, \bibinfo{author}{{Taylor}, A.R.},
  \bibinfo{author}{{Nelson}, R.F.}, \bibinfo{year}{1989}.
\newblock \bibinfo{title}{{X-ray and radio observations of flares from the RS
  Canum Venaticorum system UX Arietis.}}
\newblock \bibinfo{journal}{PASJ} \bibinfo{volume}{41}, \bibinfo{pages}{679}.
%Type = Article
\bibitem[{{Veronig} et~al.(2002a){Veronig}, {Temmer}, {Hanslmeier},
  {Messerotti}, {Otruba} and {Moretti}}]{Veronig:2002}
\bibinfo{author}{{Veronig}, A.}, \bibinfo{author}{{Temmer}, M.},
  \bibinfo{author}{{Hanslmeier}, A.}, \bibinfo{author}{{Messerotti}, M.},
  \bibinfo{author}{{Otruba}, W.}, \bibinfo{author}{{Moretti}, P.F.},
  \bibinfo{year}{2002}a.
\newblock \bibinfo{title}{{Temporal characteristics of solar soft X-ray and
  H$\alpha$ flares}}.
\newblock \bibinfo{journal}{Proceedings of the Second Solar Cycle and Space
  Weather Euroconference} \bibinfo{volume}{477}, \bibinfo{pages}{187}.
%Type = Article
\bibitem[{{Veronig} et~al.(2002b){Veronig}, {Vršnak}, {Temmer} and
  {Hanslmeier}}]{Veronig:2002b}
\bibinfo{author}{{Veronig}, A.}, \bibinfo{author}{{Vršnak}, B.},
  \bibinfo{author}{{Temmer}, M.}, \bibinfo{author}{{Hanslmeier}, A.},
  \bibinfo{year}{2002}b.
\newblock \bibinfo{title}{{Relative timing of solar flares observed at
  different wavelengths}}.
\newblock \bibinfo{journal}{Sol. Phys.} \bibinfo{volume}{208},
  \bibinfo{pages}{297}.
%Type = Article
\bibitem[{{Webb} et~al.(2020){Webb}, {Coriat}, {Traulsen}, {Ballet}, {Motch},
  {Carrera}, {Koliopanos}, {Authier} et~al.}]{Webb:2020}
\bibinfo{author}{{Webb}, N.A.}, \bibinfo{author}{{Coriat}, M.},
  \bibinfo{author}{{Traulsen}, I.}, \bibinfo{author}{{Ballet}, J.},
  \bibinfo{author}{{Motch}, C.}, \bibinfo{author}{{Carrera}, F.J.},
  \bibinfo{author}{{Koliopanos}, F.}, \bibinfo{author}{{Authier}, J.}, et~al.,
  \bibinfo{year}{2020}.
\newblock \bibinfo{title}{{The XMM-Newton serendipitous survey. IX. The fourth
  XMM-Newton serendipitous source catalogue}}.
\newblock \bibinfo{journal}{A\&A} \bibinfo{volume}{641}, \bibinfo{pages}{136}.

\end{thebibliography}
%% \bibliography{example_test}



%% For example:

%% \bibitem[Aladro et al.(2015)]{Aladro15} Aladro, R., Martín, S., Riquelme, D., et al. 2015, \aas, 579, A101

%%\end{thebibliography}

\end{document}